\documentstyle[12pt]{article}

\def\a{\alpha}
\def\b{\beta}
\def\c{\chi}
\def\d{\delta}
\def\e{\epsilon}		
\def\f{\phi}			
\def\g{\gamma}
\def\h{\eta}

\def\j{\psi}
\def\k{\kappa}			
\def\l{\lambda}
\def\m{\mu}
\def\n{\nu}
\def\o{\omega}
\def\p{\pi}			
\def\r{\rho}			
\def\s{\sigma}			

\def\D{\Delta}
\def\F{\Phi}
\def\J{\Psi}
\def\L{\Lambda}
\def\O{\Omega}
\def\P{\Pi}

\def\sfG{{\sf G}}			
\def\sfH{{\sf H}}

\def\ca{{\cal A}}
\def\cb{{\cal B}}
\def\cc{{\cal C}}
\def\cd{{\cal D}}
\def\ce{{\cal E}}
\def\cf{{\cal F}}
\def\cg{{\cal G}}
\def\ch{{\cal H}}

\def\co{{\cal O}}

\def\bd{\begin{displaymath}}
\def\ed{\end{displaymath}}
\def\be{\begin{equation}}
\def\ee{\end{equation}}
\def\bq{\begin{eqnarray}}
\def\eq{\end{eqnarray}}
\def\bqn{\begin{eqnarray*}}
\def\eqn{\end{eqnarray*}}

\def\half{\frac{1}{2}}

\def\lbl{\label}
\def\dd{\partial}
\def\der{\dd_{+}}
\def\dir{\dd_{-}}
\def\rar{\rightarrow}
\begin{document}
\setcounter{page}{36}

\renewcommand{\theequation}{\arabic{section}.\arabic{equation}}
\newcommand{\sect}[1]{\section{#1}\setcounter{equation}{0}}
\newcommand{\ad}[1]{#1^{\dagger}}
\newcommand{\lp}{\left(}
\newcommand{\rp}{\right)}
\newcommand{\nit}{\noindent}
\newcommand{\ct}[1]{\cite{#1}}
\newcommand{\bi}[1]{\bibitem{#1}}
\newcommand{\dg}{\dagger}

\title{On the Gupta-Bleuler Quantization of the Hamiltonian
Systems with Anomalies}
\author{
J. Kowalski-Glikman\thanks{e-mail address: t18@ nikhef.nl}
 \\ Center for High Energy Astrophysics\\ P.O. Box 41882,\\ 1009 DB Amsterdam,
The
Netherlands}
\maketitle
\begin{abstract}
The Gupta-Bleuler quantization method of QED can be generalized to
canonically quantized constrained systems  with quantum second-class
constraints. Such constraints may originate either from the second-class
constraints, already presented in the classical description of the theory,
or may have their sources in  quantum effects, in which case the theory
is called anomalous. In this paper, I present a detailed description of
how the Gupta-Bleuler ideas can be implemented in these cases
and I argue that there are in principle  no inconsistencies in quantum
anomalous theories. Having quantized the anomalous theories
canonically, I derive the path integral formulation
of such theories and show that some new terms are necessarily present
in this formulation. As an example, I show how the chiral Schwinger model
can be quantized in the original fermionic formulation with no
reference to the bosonized version used in the literature so far.
\end{abstract}

\sect{Introduction}
The quantum anomaly \ct{an} is the breakdown of  classical symmetries
of a system caused by
quantum effects. Even if the anomalies in global symmetries provide us
with understanding of an important class of  physical effects like
$\p \rightarrow 2\g$ decay, the anomalies of local (gauge)  symmetries
of a theory have been for a long time   regarded as an incurable disease.
It was often argued that such anomalies make it impossible to  use the Ward
identities, which may result in the lack of the proof of renormalizibility,
Lorentz invariance, and unitarity. For this reason,   one of the
most often invoked requirements for a consistent quantum theory is
that the local symmetries of the system must be free from
anomalies. This requirement has been successfully applied e.g.,
to restrict the representation content under the gauge group of fermions
in the standard model and to derive the critical dimension in string
theory.

Only  few years ago,  the applicability of the standard
quantization procedures to anomalous theories has been put in
question  \ct{Fd}.  It was observed that one should not use  information
concerning the classical symmetry directly in the quantum context.
The very reason for that are the anomalies themselves -- the symmetry
group of the quantum theory may be different from the corresponding
classical ones, which is not very much surprising after all. Therefore,
one may conclude that the problems do not reside in the anomalous theories
themselves but rather in an inappropriate way of quantizing them.
The problem will become even more severe if one realizes that
the anomaly may also provide an enlargement of the symmetry group
of quantum theory\footnote{This effect, called antianomaly is briefly discussed
in the last section of this paper},
 in which case all standard techniques of quantization
(like the Dirac approach or standard path integral method) cannot be directly
applied.

It seems clear that the only way out of these problems is to discuss
the gauge symmetries
of a theory not prior  but after quantization. Since in
the language of canonical quantization anomalies exhibit
themselves as the  non -- closure of the commutator algebra of
classical symmetry generators, this would amount in developing
the quantization scheme such that one can deal
with a general system of first- and second-class constraints
on the quantum level.
Actually such a scheme is well known for almost 40 years and has its
roots in the Gupta-Bleuler approach to the quantization of
quantum electrodynamics \ct{GB}:

If one wants to implement the covariant gauge-fixing condition
of QED $\dd_{\m}A^{\m} = 0$
directly on states of quantum theory, one encounters an immediate
difficulty realizing that the negative and positive frequency
parts of this condition, $\dd_{\m}A^{(+)\m}=0$ and $\dd_{\m}A^{(-)\m}=0$
do not commute and therefore there are no nontrivial
states satisfying the condition
\be
\dd_{\m}A^{\m}|phys> = 0.   \lbl{ii1}
\ee

\nit
The simple solution to this problem is to realize that in quantum theory
the only physically meaningful objects are the matrix elements of
quantum operators
and therefore, instead of (\ref{ii1}), one should  use
\be
<phys|\dd_{\m}A^{\m}|phys'> = 0,   \lbl{i2}
\ee

\nit
to define gauge-fixed physical state.
The idea of Gupta and Bleuler was to consider a particular necessary condition
for
(\ref{i2}) given by
\be
\dd_{\m}A^{(+)\m}|phys> = 0.
\lbl{i3}
\ee

\nit
It was proved that  condition (\ref{i3}) describes the
physical states of QED correctly.

This method can be easily generalized to the canonical Hamiltonian systems
with quantum second-class constraints. To this end, one splits
the system of real (Hermitean) constraint operators $\cg$ whose algebra
does not close
\be
[\cg_{A}, \cg_{B}] = \o_{AB}
\ee

\nit
into holomorphic and antiholomorphic parts $\cg_{I}$ and $\ad{\cg}_{I}$,
both of whose have separately closed algebras
and defines physical states as to satisfy
\be
\cg_{I}|phys> = 0.
\ee

\nit
It should be stressed that in this way we effectively treat the holomorphic
part of the second-class constraints as generators of a gauge symmetry,
which leaves the physical states invariant.

In this paper, I describe the Gupta-Bleuler procedure applied to
quantization of theories with second-class constraints and/or
anomalies in general setting. After reviewing the standard Dirac procedure
in Section 2, in two subsequent sections I develop the general Gupta-Bleuler
approach. In Section 5, I discuss a possible formulation of the Gupta-Bleuler
quantization in terms of the path integral technique. In Section 6,
I discuss the Gupta-Bleuler quantization of the simple anomalous theory,
the chiral Schwinger model, which in complete agreement with earlier
works is shown to be finite and soluble and is equivalent to the
quantum theory of a massive and massless scalar fields. Finally,
Section 7 is devoted to discussion of the so -- called
antianomaly effect and Section 8 contains conclusions and some remarks.

\sect{Review of  Dirac theory of quantization of constrained systems}

In this section, I  give a short review of  Dirac theory of
constrained systems \ct{Dir1}. There are many   textbooks
in which the extensive discussion of this subject can be
found:  \ct{Dir}, \ct{HRT}, \ct{Sund} (see also  review papers  \ct {Sen},
\ct{Mar});
in what follows, I will mainly refer to the new  excellent book
by Henneaux and Teitelboim \ct{HT} covering
also the most recent development on the field.
Therefore, below I will concentrate only on the aspects
of the problem which will be relevant for the theory developed later.

Let the system of interest be described by the Lagrangian $L$, which
depends on positions $q_{I}$ and velocities $\dot{q}_{I}$, where
the index $I$ can be discrete (mechanical systems) or continuous
(field theory):
\be
L = L(q_{I}, \dot{q}_{I}).
\ee

\nit
One defines momenta in the standard way,
\be
p_{I} =  \frac{\dd L}{\dd \dot{q}_{I}}(q, \dot{q}).   \lbl{mom}
\ee

\nit
It may happen that
the Hessian matrix $W_{IJ} = \frac{\dd p_{I}}{\dd \dot{q}_{J}}
= \frac{\dd^{2} L}{\dd \dot{q}_{I}\dot{q}_{J}}$
does not have a maximal rank and therefore
not all of the equations (\ref{mom}) can be solved
for the velocities. Instead, we have some number of relations
\be
\cg_{A} = p_{A} - \j_{A}(q_{I}, p_{I}) = 0, \lbl{cons}
\ee

\nit
where  the maximal number of velocities
have been expressed in terms of momenta using
(\ref{mom}). These relations are called (primary) constraints, since they
restrict
the possible dynamics of the system to the surface in the full phase
space of the system defined by eq. (\ref{cons}).

Given the symplectic structure on the phase space of the problem
(i.e., the Poisson bracket $\{\star , \star \}$), we can calculate the
bracket of constraints $\cg_{A}$ to be in general
\be
\{\cg_{A}, \cg_{B} \} = f_{AB}^{\;\;\;C}\cg_{C} + \o_{AB}, \lbl{ca}
\ee

\nit
where $f_{AB}^{\;\;\;C}$ and $\o_{AB}$ may be phase space point-dependent,
and $\o_{AB}$ is supposed to be non -- vanishing identically on the constraints
surface
(\ref{cons}).
We assume  that $\o_{AB}$ is such that the Jacobi identity
for an bracket algebra (\ref{ca}) holds.
Among  the constraints, there are such  that their
bracket algebra closes,
\be
\{\cg_{\a}, \cg_{\b} \} = f_{\a\b}^{\;\;\;\g}\cg_{\g},  \lbl{fc}
\ee

\nit
and which have a vanishing bracket with all other constraints on the
constraints
surface ($\cg = 0$). In Dirac terminology such constraints are called (primary)
first-class.
All other constraints are called second-class and  will be denoted
 by $\cg_{i}$.

{}From the Lagrangian of a theory we can construct the canonical Hamiltonian
by performing
Legendre transformation
\be
H_{0}(p,q) = p_{I}\dot{q}_{I} - L(q, \dot{q}),   \lbl{h0}
\ee

\nit
where on the right hand side
 we replaced all velocities by positions and momenta wherever
possible or use the relations (\ref{cons}) otherwise. One can
easily check explicitly that the canonical Hamiltonian defined above
is velocity-independent.

Since the dynamics is supposed to take place on the constraints surface, we
can add to the canonical Hamiltonian a general linear combination of
constraints and form in this way the so-called primary Hamiltonian
\be
H = H_{0} + \l^{\a}\cg_{\a} + v^{i}\cg_{i},     \lbl{ham}
\ee

\nit
which is a generator of dynamics. The important consistency
requirements is that the
time evolution takes place on the
constraints surface, which means that the constraints are
supposed to be conserved in time,
to wit
\be
\frac{d\cg_{A}}{dt} = \{\cg_{A}, H\} = V^{B}_{A}\cg_{A}.
\lbl{ddtcon}
\ee

\nit
It may happen however that the set of primary constraints does not
satisfy the above equation. Except for uninteresting case when this relation
becomes inconsistent (of the form 0 = 1), it is  possible that
some new relations between phase space variables
emerge. These new relations are called secondary constraints and should be
added to the set of primary ones\footnote{It is clear that there is at most
as many
secondary constraints as the first-class primary ones (keep in mind
however that the first-class primary constraints may become second-class
in the full theory i.e., when all secondary, tertiary, etc. constraints
are found as described below).
Indeed, if $\cg_{i}$ is second-class then
\be
0 = \{ H, \cg_{i} \} = \{ H_{0}, \cg_{i} \} + v_{j}\o_{ji}, \nonumber
\ee
where the last equality holds up to some combination of constraints.
This equation is always solvable, since $\o$ is invertible by definition.}.
Then the procedure of
splitting the constraints into first and second-class and looking for new
ones should be repeated along the same line of reasoning,
until all  constraints are found, i.e., until eqn. (\ref{ddtcon}) is satisfied
identically.

The important observation is that the first-class
constraints are in one-to-one correspondence with the (local, i.e., with time
-- dependent parameter)
gauge symmetries
of the
theory and (\ref{fc}) is in fact a representation of the gauge group.
The same is true after quantization: actually the whole
idea behind the Dirac quantization was to ensure the gauge invariance
of quantum theory. In the Dirac approach, the gauge invariance  is introduced
by demanding that the operators corresponding to the gauge
generators annihilate the wave function. The anomaly arises when these
operators
fail to have closed commutators, which means that the classical symmetries are
broken on quantum level. This important facts are discussed in details
in \ct{HT}.

Given a full set of constraints $\cg_{\a}$, $\cg_{i}$, we can turn to the
quantization problem. Before doing that, we introduce a concept of
Dirac bracket which replaces the standard Poisson bracket in the Dirac
quantization procedure.

The second-class constraints  satisfy a simple Poisson bracket algebra
on the constraint surface
\be
\{\cg_{i}, \cg_{j} \}|_{\cg =0} = \o_{ij},
\ee

\nit
where $\o_{ij}$ is invertible. Then one defines the Dirac bracket
of any two functions on a constraint surface to be
\be
\{\ca , \cb\}^{\ast} = \{\ca , \cb\} - \{\ca , \cg_{i}\}(\o^{-1})^{ij}
\{\cg_{j}, \cb\}.
\ee

This bracket can be understood as a projection of the Poisson bracket
down to the second-class constraint surface.
Thus, we can consistently replace all the Poisson brackets by Dirac
brackets  and  solve the second-class
constraints, expressing all $p_{i}$ by $\j_{i}$ using (\ref{cons}).
The Dirac bracket provide us with a new symplectic
structure on the phase space of the problem, such that the second-class
constraints are absent.
The quantization is achieved by replacing the Dirac brackets
by  commutators in the standard way and finding the representation
of the canonical commutational relations on some Hilbert space.

As for the first-class constraints,
the Dirac proposal consists of implementing them directly
on the quantum level. After quantization, the first-class constraints
become Hermitean operators satisfying the commutator algebra (provided there
are no
anomalies)
\be
[\cg_{\a}, \cg_{\b}] = if_{\a\b}^{\;\;\;\;\g}\cg_{\g}, \lbl{ccon}
\ee

\nit
which is in direct correspondence with the Poisson (Dirac) bracket algebra
(\ref{fc}). In the formula above $f_{\a\b}^{\;\;\;\;\g}$ are, in
general, operator - valued and, of course, one may encounter the notorious
operator ordering problem in passing from (\ref{fc}) to (\ref{ccon}).

Given the operators corresponding to the first-class constraints, one
implements them by demanding that the quantum dynamics takes place
on the subspace of an original Hilbert space, called the physical
Hilbert space, consisting of all vectors (wave functions),
which are annihilated by the constraints, to wit
\be
\cg_{\a}|phys> = 0. \lbl{phys}
\ee

\nit
As it was said before, since the first-class constraints correspond to the
gauge symmetries of the
theory, this condition means simply that the physical wave functions are gauge
invariant.

It follows from the construction discussed above  that since the Hamiltonian
operator constructed from the primary Hamiltonian commutes with constraint
operators on physical states, the time evolution maps physical Hilbert space
into itself. Analogously, we define  physical observables to be Hermitean
operators commuting with constraints. In a final point of
construction we define an inner product on the physical subspace to be
the one induced from the inner product of an initial Hilbert space of the
problem. One should keep in mind that very often the initial  space
of states does not have a positive definite inner product (so, strictly
speaking it is incorrect to call this space a Hilbert space) and only
in correctly constructed theory there are no negative and zero norm states
in physical sector (if there are negative norm states in the physical
sector one may try to construct a second quantized theory in the standard way).
\newline

Let us conclude this section by discussing the Dirac quantization
of a   massive scalar field theory on $S^{1}\times R^{1}$ in light cone
coordinates\footnote{This example will be relevant to what follows. It happens
that it describes the effective theory of 2 -- dimensional chiral QED
(Schwinger model)
-- see Section 6}.
The Lagrangian of the model reads
\be
L(x^{+}) = \frac{1}{2\p}\int_{0}^{2\p} (\dd_{+}\F\dd_{-}\F - \half
m^{2}\F^{2}).
\ee

\nit
Expanding into Fourier modes in $x^{-}$
\be
\F = \sum_{n}\F_{n}e^{inx^{-}}, \hspace{.5cm} \F_{-n} = \ad{\F}_{n},
\ee

\nit
we have
\be
L(x^{+}) = \sum_{n>0}(-in\dot{\F}_{n}\ad{\F}_{n}  + in\dot{\F}^{\dg}_{n}\F_{n}
- m^{2}\F_{n}\ad{\F}_{n}) - \half m^{2}\F_{0}^{2},
\ee

\nit
where the overdot denotes  differentiation with respect to the
light cone time $x^{+}$.
One can readily find the momenta
\bq
P_{n} &=& \frac{\dd L}{\dd\dot{\F}_{n}} = -in\ad{\F}_{n},  \\
\ad{P}_{n} &=& \frac{\dd L}{\dd\dot{\F}^{\dg}_{n}} = in\F_{n} , \\
P_{0} &=& 0.
\eq

\nit
We see therefore that there are constraints in the theory. The first two of
them are clearly second-class.
The canonical Hamiltonian reads
\be
H = \sum _{n>0}m^{2}\F_{n}\ad{\F}_{n} + \half m^{2}\F_{0}^{2}.
\ee

\nit
We see that there is a secondary constraints $\F_{0} = 0$, which together
with $P_{0}= 0$ form a second class system.
We can get rid of them by using the Dirac bracket
and erase them from the all relevant formulae. Since the
non -- zero `momentum' constraints (labelled by non -- zero index $n$)
do not depend on $\F_{0}, P_{0}$,
we need only to modify the Hamiltonian, to wit
\be
H = \sum _{n>0}m^{2}\F_{n}\ad{\F}_{n}.
\ee

\nit
Let us turn to the remaining constraints
\bq
\cg_{n} &=& \ad{P}_{n} - in\F_{n}, \\
\ad{\cg}_{n} &=& P_{n} + in\ad{\F}_{n},
\eq

\nit
which satisfy the Poisson bracket algebra
\be
\{\cg_{n}, \ad{\cg}_{m}\} = -2in\d_{n,m}.
\ee

\nit
Now we can write down  the Dirac bracket
\be
\{\ca , \cb\}^{\ast} = \{\ca , \cb\} + \frac{i}{2n}\{\ca , \cg_{n}\}
\{\ad{\cg}_{n}, \cb\} - \frac{i}{2n}\{\ca , \ad{\cg}_{n}\}
\{\cg_{n}, \cb\}
\ee

\nit
use this bracket and replace $\ad{\F}, \ad{P}$
by $\F$ , $P$
  everywhere using the constraint equations.
The fundamental bracket is now
\be
\{P_{n}, \F_{m}\}^{\ast} = -\half\d_{n,m}
\ee

\nit
and the commutator  (after rescaling $\F \rightarrow \half\F$)
\be
[P_{n}, \F_{m}]   = -i\d_{n,m}.
\ee

\nit
Substituting the constraint (2.30) into Hamiltonian we obtain
\be
H = \sum _{n>0}m^{2}\frac{i}{2n}\F_{n}P_{n},
\ee

\nit
up to the ordering ambiguity. In the coordinate representation
\be
P_{n} = -i\frac{\dd}{\dd \F_{n}},
\ee

\nit
and we have the Schr\"odinger equation
\be
\sum _{n>0}m^{2}\frac{1}{2n}\F_{n}\frac{\dd}{\dd \F_{n}}\J (\F_{n}) =
\ce_{\J}\J (\F_{n})
\label{s1}
\ee

\nit
Before solving this equation let us discuss the form of the Hilbert space
of wave functions.
In this space the constraint equations must be satisfied on the
operator level i.e.,\footnote{This shows that the second-class constraints
are not totally `forgotten' by the quantum theory even if the phase space
of the problem is reduced to the constraint surface with a symplectic
structure given by the Dirac bracket.}
\be
\ad{P}_{n} = i\F_{n},
\label{her}
\ee

\nit
where we already rescaled the operators $P_{n} \rightarrow
\frac{1}{\sqrt{n}}P_{n},
\F_{n} \rightarrow \sqrt{n}\F_{n}$. This rescaling does not alter
the canonical commutational relations and the form of the Hamiltonian operator.
Let us assume that the Hilbert space of the problem is a space of polynomials
in  $\F_{n}$ variables. This assumption follows from the fact that the
solutions of (\ref{s1}) are polynomials. For the basic elements of such
space, the monomials $\F_{n}^{\a}$, we define the inner product to be
\be
<a\F_{n}^{\a}, b\F_{m}^{\b}> = a^{\ast}b(\b !)\d_{n,m}\d^{\a ,\b}.
\ee

\nit
It is easy to see that this inner product is consistent with
hermicity relations (\ref{her}).

Now,  equation (\ref{s1})
is solvable in terms of linear combination of products of
finite number of monomials
\be
\J = \prod\F^{\a_{n}}_{n}.
\ee

\nit
The energy of such solution is
\be
\ce_{\J} = \sum \frac{m^{2}}{2n}\a_{n}.
\ee

\nit
Therefore, we have solved the quantum theory of a scalar field using
Dirac method. Note that the second-class constraints played an important
role even after Dirac
bracket has been introduced and replaced by the commutator. Indeed they
provide us with a hint as to how define the Hilbert space of the problem
and make the Hamiltonian operator Hermitean. One should also observe that the
Schr\"odinger equation (\ref{s1}) does not have any non -- trivial solution
in the space of square integrable functions\footnote{It is possible to solve
the theory in the Hilbert space being a space of square integrable
functions. In order to do that, one needs to change the representation of
commutational
relations (2.34) and include an exponential dumping factor in the
wave function. Then the result is identical to the one presented in
Section 4.}.

\sect{Gupta-Bleuler constraints}
Even though the Dirac procedure of quantization of  constraint systems
seems simple in principle, one may encounter many problems in applications
to particular models. First of all, the required split of constraints
into first and second-class may be problematic if one wants to make it
covariant with respect to some global symmetries of the theory like
a global target space super Poincare symmetry in superparticle and
superstrings models. Secondly, it may happen that the algebra
of the commutators of constraints (\ref{ccon}) acquires additional
central charge terms, as compared to the Poisson (Dirac) bracket algebra
(\ref{fc}). This purely quantum effect is in fact the exhibition
of anomalies discovered originally in the diagrammatic context
in late sixties.

The appearance of anomalous terms in (\ref{ccon})
makes it impossible to implement the Dirac procedure
of recognizing the physical subspace of the Hilbert space as
the vector space of states annihilated by constraints. Indeed,
if the system of
operators $\cg_{\a}$ corresponding to first-class constraints
satisfy the anomalous commutator algebra\footnote{Strictly speaking,
for anomalous theories, the commutator below is defined only
with respect to some pre -- chosen set of states (i.e., one should
use the `sandwiched' form of the algebra). Since we always (implicitly)
define the Hilbert space  as the first step of quantization procedure,
in what follows we will ignore this subtle point.}

\be
[\cg_{\a}, \;\cg_{\b}] = if_{\a\b}^{\;\;\;\g}\cg_{\g} + \omega_{\a\b},
\label{rcon}
\ee

\nit
instead of (\ref{ccon}), and
we want  to follow the Dirac procedure to define the physical states
as the ones satisfying (\ref{phys}),
we find immediately that

\be
\omega_{\a\b}|phys> = 0 \hspace{.2cm} \Rightarrow |phys> = 0,
\ee

\nit
since $\omega_{\a\b}$ does not have non -- trivial zero modes by assumption.
Therefore, the Dirac procedure cannot be consistently applied
to quantization of anomalous theories and requires  modifications.
In this sections, I  discuss a proposal of replacing the
Dirac procedure by a new one having its roots in Gupta-Bleuler
quantization method in quantum electrodynamics.

There are clearly many methods of imposing constraints in quantum theory --
one needs to restrict a Hilbert space by some reasonable analogue of the
classical condition of vanishing of some combination of phase-space variables.
Among these possibilities, the Dirac requirement that the constraints
annihilate the physical states, (\ref{phys})
is probably the most restrictive. As we have shown above,
it does not work for the system
of mixed constraints, and we must get rid of the second-class
ones on the classical level by making use of the Dirac brackets. The other,
physically  more appealing possibility is to define
physical states as the maximal subset of all states such that
the matrix elements of all operators corresponding to constraints
vanish between any two of
them, to wit\footnote{It is implicitly assumed that the Hilbert space of the
problem does not include any zero-norm states $|\j_0>\in \ch_0$, $\|\,|\j_0
>\|^2=0$. If such states
were present initially, they should divided out: $\ch \rightarrow \ch/\ch_0$.}

\be
<phys|\cg_{A}|phys'> = 0.
\label{Pcon}
\ee

\nit
Clearly, in this case the constraints operators $\cg_{A}$ can form
the algebra
\be
[\cg_{A}, \;\cg_{B}] = if_{AB}^{\;\;\;C}\cg_{C} + \omega_{AB},
\label{rc}
\ee

\nit
 with the non -- vanishing central charge
$\omega_{AB}$ which corresponds either to the classical
second-class constraints, or has its source in anomalies,
without breaking the consistency. However,
condition (\ref{Pcon}) is not very  restrictive and one has
problems with interpretation of
physical states defined in this way (cf. discussion in \ct{Jur}).

In this paper, we choose a middle of the road method. Suppose that we can
split
the full algebra of Hermitean  constraints  $\cg_{I}$ into two subsets
consisting of
the complex constraints $\cg_{I}$ and their Hermitean conjugate
$\ad{\cg}_{I}$ such that the algebras of holomorphic and antiholomorphic
constraints close

\be
\left[ \cg_{I}, \cg_{J} \right] = if_{IJ}^{\:\:\:\:K} \: \cg_{K},
\label{Galg}
\ee

\be
\left[ \ad{\cg}_{I}, \ad{\cg}_{J} \right] = -i\ad{\cg}_{K} \:
\bar{f}^{K}_{\:\:IJ},
\label{Gadj}
\ee

\be
\left[ \cg_{I}, \ad{\cg}_{J} \right] = 2 \: Z_{IJ}.
\label{Z}
\ee

\nit
Then, one defines the physical states as a subset of the solutions of
(\ref{Pcon})
defined as

\be
\cg_{I}|phys> = 0.
\label{GBcon}
\ee

\nit
Clearly, if (\ref{GBcon}) is satisfied, the condition (\ref{Pcon}) is satisfied
as well -- indeed it follows from (\ref{GBcon}) that the matrix elements
between physical states of both $\cg_{I}$ and $\ad{\cg}_{I}$ are equal to zero.

The important question arises,  as to whether the
Gupta-Bleuler algebra of constraints (\ref{Galg} - \ref{Z}) is equivalent
to the original algebra commutator (\ref{rc}).
Before giving the detailed proof of the equivalence of these two
presentations let us describe how (\ref{Galg} - \ref{Z})
can be  constructed from (\ref{rc})\footnote{This question has been
addressed before in \ct{ZJJJ}.}.

The procedure consists of two major steps: the disentanglement of the
first-class
subalgebra and the polarization of the remaining subalgebra of (\ref{rc}).
Let us stress already at the beginning that both steps may in general cause
breakdown of some global symmetries present in the theory and
that some of resulting operators may be non -- local.

In the first step,  we construct the {\em maximal} set of vectors $v^{A}_{\a}$
labelled by the index $\a$, such that $v$ are built out of Hermitean operators
and they commute both with $\cg_{A}$ and $\o_{AB}$. Suppose the
following condition is satisfied:
\be
v^{A}_{\a}\o_{AB} = 0.   \label{v}
\ee

\nit
Then it is easy to see that $\cg_{\a} = v^{A}_{\a}\cg_{A}$ are first-class
i.e., they form the closed algebra  with all constraints. Indeed
\bq
[\cg_{\a}, \cg_{A}] &=& [v^{B}_{\a}\cg_{B}, \cg_{A}] =\nonumber \\
&&v^{B}_{\a}\o_{BA} + \mbox{terms proportional to }\cg_{A},
\eq

\nit
and from (\ref{v}) the first term vanishes. Observe that by  construction
$\cg_{\a}$ are Hermitean.

Now we split the whole set of constraints $(\cg_{A})$ into two subalgebras,
one formed by $\cg_{\a}$ and defined above and the second consisting of
remaining
operators, which we will call $\cg_{i}$,
which satisfy
\be
[\cg_{i}, \cg_{j}] = if_{ij}^{\:\:\:A}\cg_{A} + \o_{ij},
\ee

\nit
where now the operator-valued matrix $\o_{ij}$ has a maximal rank. The
polarization
of the second-class constraints $\cg_{i}$ amounts  in constructing  an
operator-valued matrix $J_{ij}$ (a complex structure), which commutes
with $\cg_{i}$ and satisfies
\bq
J_{ij}J_{jk} = -\d_{ik}, \\
J_{ij}\o_{jk} - J_{kj}\o_{ji} = 0.
\eq

\nit
It is easy to see that once these conditions hold, the complex constraints
$\tilde{\cg}_{i} = \cg_{i} + iJ_{ij}\cg_{j}$ and
$\tilde{\cg}^{\dg}_{i} = \cg_{i} - iJ_{ij}\cg_{j}$ satisfy the
algebra of the form  (\ref{Galg} - \ref{Z}). In the final step we form a full
set of generators of the Gupta-Bleuler constraints algebra by taking
$\cg_{I} = (\cg_{\a},  \tilde{\cg}_{i})$ and
$\cg^{\dg}_{I} = (\cg_{\a},  \tilde{\cg}^{\dg}_{i})$.

One should observe that
the constraint algebra can be transformed into even simpler algebra, namely
\be
[\sfG_{\a}, \sfG_{\b}] = [\sfG_{\a}, \sfG_{i}] = [\sfG_{i}, \sfG_{j}] = 0,
\lbl{ab1}
\ee
\be
[\sfG_{i}, \ad{\sfG}_{j}] = i\d_{i,j}
\lbl{ab2}
\ee

\nit
which I will call `abelian Gupta-Bleuler form'
(here and below I denote constraints in abelian form by $\sfG$). It is clear
from this form of the algebra that the holomorphic and antiholomorphic
constraints  essentially behave like annihilation and creation operators.
Indeed, it follows from (\ref{ab1}), (\ref{ab2})
that there exist
a local coordinate system $({\bf P}, {\bf Q}; p_{\a}, q_{\a};
p_i, q_i)$ in vicinity of the constraint surface such that there exist
a symplectomorphism $\sfG_{\a}\rar p_{\a}$, $\sfG_i\rar \half(p_i+iq_i)$
$\ad{\sfG}_i\rar \half(p_i-iq_i)$. Here ${\bf P}$ and ${\bf Q}$ are
coordinates `orthogonal' to the constraints surface.

As we saw above, the problem of bringing an algebra of constraints
to the (abelian) Gupta-Bleuler form is a problem of solving systems
of differential equations for $v$ and $J$. Using Darboux theorem, it is
possible to
demonstrate  (cf. \ct{HT} and the detailed discussion
in \ct{WK}) that these equations are always solvable locally\footnote{The only
exception is a system consisting of odd number of fermionic
constraints. Subsystems are pretty rare and, besides, one can applied the Gupta
-- Bleuler theory
to any subsytem of even number of constrains and solve the remaining ones
by the standard Dirac method. Moreover, since our main goal to
construct the consistent quantum theory with anomalies, such problems
never arise (systems with odd number of fermionic constraints
are quantum mechanical not field theoretical).}. One should
stress however that the proof is essentially classical in the sense
that it make use of notions of phase space, constrained surface, and
Poisson bracket. Therefore, one should be very careful about
resolving the ordering problem at the beginning of the procedure
and tracking any ordering ambiguity which may arise so that
the correspondence between commutators and Poisson brackets is kept.

On the other hand, it should be observed that the proof merely guarantees
solvability of the polarization problem and is not operational in general
(in the sense that
it does not prescribe the step -- by -- step procedure). This procedure
is either obvious (as in the case of examples considered in this paper) or
very difficult (as in the case of anomalous Gauss law algebra of
chiral QCD, see \ct{DT}). In the second case, it is little hope that
the canonical approach is capable of producing anything more than a
perturbation
theory; but then one can equally well turn to the path integral approach,
which, as it will be shown below, does not require explicit Gupta -- Bleuler
polarization, but only its existence.

Therefore, the conclusion is that the Gupta -- Bleuler procedure
is feasible in all cases of interest.

\sect{The Gupta-Bleuler Quantization Method}
Let me now describe the Gupta-Bleuler quantization method
for theories with second-class constraints and/or anomalies. We
start with a classical Lagrangian $L$. From it we deduce a phase space
with its symplectic structure,
the set of primary real constraints $\cg_{A}^{0}$ constraints and the canonical
Hamiltonian $H_{0}$. Then we quantize: we define a representation
of canonical commutational relations
\be
[p_{I}, q_{J}] = -i\d_{IJ}
\ee

\nit
on some Hilbert space (again, strictly speaking it will be an inner product
space since some vectors may have negative norm). From the algebra of
Hermitean constraints,
\be
[\cg^{0}_{A}, \cg^{0}_{B}] = if_{AB}^{\:\:\:C}\cg^{0}_{C} + \o_{AB},
\ee

\nit
we extract the first-class ones, $\cg_{\a}$, polarize the remaining ones,
and obtain in this way two sets of holomorphic $\cg^{0}_{I}$ and
antiholomorphic $\cg^{0\dg}_{I}$
constraints, where both of them include the first-class ones. In this way
the algebra of constraints is cast to the Gupta-Bleuler form
(\ref{Galg} - \ref{Z})
We
define primary physical states $|phys,0>$ to be annihilated by
the primary holomorphic constraints, to wit
\be
\cg^{0}_{I}|phys,0> = 0 \lbl{pph}
\ee

\nit
It is clear that in theories with anomalies it would be inappropriate
to look for secondary constraints already on classical level.
Indeed, as was noted in footnote 2, the secondary constraints
results from primary first-class ones and since the number of them
may change due to the quantum effects the number and form of
secondary constraints may change also.

We start the construction of secondary constraints with definition
of primary Hamiltonian operator $H$, which by assumption generates a time
evolution by virtue of Schr\"odinger equation,
\be
H = H_{0} + u^{I}\cg_{I}^{0} + \cg^{0\dg}_{I}v^{I}.  \lbl{hami}
\ee

\nit
Observe that $H$ is Hermitean between primary physical states.
In complete analogy with Dirac approach, we assume that the time
evolution maps physical states into themselves, and therefore
the primary Hamiltonian (\ref{hami}) must commute with holomorphic constraints
on physical states. We have therefore an equation
\be
0 = [\cg^{0}_{I}, H]|phys,0> = ([\cg^{0}_{I}, H_{0}]
  + Z^{0}_{IJ}v^{J} + \cg^{0\dg}_{J}[\cg_{I}^{0}, v^{J}])|phys, 0>,
\lbl{newcon1}
\ee

\nit
which after little algebra can be rewritten as
\be
(\cg^{0}_{I}H_{0} + \cg^{0}_{I}\cg^{0\dg}_{J}v^{J})|phys, 0> = 0.
\lbl{newcon}
\ee

\nit
The equations above can be either solved for $v^{J}$, or produce new
constraints
which should be included into the constraints set, polarized, if necessary
and then the procedure of hunting for new constraints should be repeated.
After the whole set of constraints of the theory has been revealed,
one can construct
the observables  in complete analogous way:
 take any Hermitean operator, add a linear combination of constraints and then
check if it commutes with holomorphic ones on physical states.
Having obtained the full set of constraints $\cg_{I}$ and $\ad{\cg}_{I}$ we
define
the physical states to be annihilated by the holomorphic ones and the time
evolution
is subject to Schr\"odinger equation
\be
(\frac{\dd}{\dd t} - H)|phys, t> = 0,
\ee

\nit
where $H$ is defined in (\ref{hami}).
This concludes the construction of quantum
theory\footnote{Let me stress here again: the idea was to take as few bits
of information from the classical theory as possible -- the primary constraints
and
the primary Hamiltonian and perform all subsequent steps on quantum level. Of
course,
one can try another approach, namely, to reveal all classical constraints and
canonical
Hamiltonian on the classical level, then quantize and try to make the resulting
theory
consistent as described above. There is little hope that these two procedures
give
the same result in general.}.

Let us now return to the example of Section 2 and analyse it again from the
point
of view of Gupta-Bleuler method.

After performing the same steps as before, we see that the
system is described by the set of holomorphic and antiholomorphic constraints
\bq
\cg_{n} &=& \ad{P}_{n} - in \F_{n},\\
\cg_{0} &=& P_{0} - im^{2}\F_{0}, \\
\ad{\cg}_{n} &=& P_{n} + in\ad{\F}_{n}, \\
\ad{\cg}_{0} &=& P_{0} + im^{2}\F_{0},
\eq

\nit
and the canonical Hamiltonian
\be
H = \sum_{n>0} m^{2}\F_{n}\ad{\F}_{n}.
\ee

\nit
The Gupta-Bleuler Hamiltonian can be obtained by using the procedure
described above and reads
\be
H_{GB} = \half m^{2}\sum_{n>0}(\F_{n}\ad{\F}_{n} + \frac{i}{n}\F_{n}P_{n}).
\label{scalham}
\ee

\nit
Let our Hilbert space be a space of square integrable functions $\J$. Solving
the
holomorphic constraints $\cg_{n}\J = \cg_{0}\J = 0$ gives
\be
\J =
e^{-\frac{m^{2}}{2}\F_{0}^{2}}e^{-\sum_{n>0}n\F_{n}\ad{\F}_{n}}\J_{1}(\F_{n}),
\label{wf}
\ee

\nit
where $\J_{1}$ is a polynomial. Substituting this result
to the Schr\"odinger equation
\be
\half m^{2}\sum_{n>0}(\F_{n}\ad{\F}_{n} + \frac{i}{n}\F_{n}P_{n}))\J = \ce\J ,
\ee

\nit
we obtain, after subtracting the infinite energy due to the ordering
change in the second term an equation for $\J_{1}$
\be
\frac{im^{2}}{2n}\F_{n}P_{n}\J = \ce\J ,
\ee

\nit
which is nothing but the equation (\ref{s1}) discussed in the Section 2.
Therefore we conclude that the Gupta-Bleuler approach gives
rise to the same energy spectrum and  Hilbert space
as the standard Dirac procedure\footnote{The only difference in the
form of the Hilbert space, the $e^{-\frac{m^{2}}{2}\F^{2}_{0}}$ factor,
is irrelevant, as the only observable which does not annihilate it is a
constant in $\F_{0}$, $P_{0}$ sector.}. One should observe that
the reason for using holomorphic constraints is that there are
no solutions to the antiholomorphic constraints equations in the space
of integrable functions. Our construction above can be readily applied
to the massless case, $m^{2} = 0$. In this case the Hamiltonian is zero
and the theory is described by the holomorphic constraints $P_{n} - in\F_{n}$
and the single first-class constraint $P_{0}$. The wave function of the
massless
scalar field is therefore
\be
\J_{massless} = e^{-\sum_{n>0}n\F_{n}\ad{\F}_{n}}\J_{1}(\F_{n}).
\ee

To show how the Gupta -- Bleuler technique work in the real field theoretical
framework, let us consider yet another example, namely the massive
four-dimensional
QED. The Lagrangian of the theory reads
$$
L=-\frac14F_{\m\n}F^{\m\n}+\half m^2A_{\m}A^{\m}=$$\be
=-\frac14(F_{ij})^2+\half (\dd_0A_i-\dd_iA_0)^2+\half m^2A_0^2-\half m^2A_i^2.
\ee

\nit
By using standard procedure, we easily find two sets of
constraints\def\pa{\partial}
$$
Q_0(x) = \pa_iP_i(x)+ m^2A_0(x)\approx 0,
$$$$
P_0(x) \approx 0
$$

\nit
and the Hamiltonian
$$
H= \int\, d^3x\lp \half P_i^2 + \half m^{-2}(\pa_iP_i)^2+\half m^2A_i^2+\frac14
F_{ij}^2+\l P_0+\r Q_0\rp.
$$

We assume the following form of canonical commutational relations
$$
[P_{\m}(x), A_{\n}(y)] = i\h_{\m\n}\d^3(x-y), \hspace{.5cm}
\h_{\m\n}=(1,-1,-1,-1).
$$

The constraints $Q_0$ and $P_0$ are clearly second class. Since $Q_0$ does not
commute
with $A_i$, we make the replacement $A_i\rar \ca_i= A_i+m^{-2}\pa_iP_0$.
It is easy to see that the change of  Hamiltonian can be absorbed into
redefinition
of $\l$ so that we have
$$
H= \int\, d^3x\lp \half P_i^2 + \half m^{-2}(\pa_iP_i)^2+\half
m^2\ca_i^2+\frac14
F_{ij}^2+\l P_0+\r Q_0\rp
$$

\nit
with
$$
F_{ij}=\pa_i\ca_j-\pa_j\ca_i,
$$$$
[P_i(x),\ca_j(y)]=-\d_{ij}\d^3(x-y),
$$$$
[P_0(x),Q_0(y)]= im^2\d^3(x-y).
$$

Dirac quantization of the theory is strightforward. One just use the Dirac
bracket
and forgets
about $P_0$ and $Q_0$ whatsoever; it remains to look for energy eigenstates
of the Hamiltonian
$$
H\J_D=\ce_{\J}\J_D.
$$

Interestingly enough, the theory is fully solvable. To see that we first
remove the $\frac14F_{ij}^2$ term from the Hamiltonian by replacing
\be
\J_D(\ca)=e^{\pm\int d^3x\e^{ijk}\ca_i\pa_j\ca_k}\j(\ca).\lbl{WZW}
\ee

The wave function $\j(\ca)$ satisfies
$$
\int\, d^3x\lp \half P_i^2 + \half m^{-2}(\pa_iP_i)^2+\half
m^2\ca_i^2\rp\j(\ca)=\ce_{\J}\j(\ca).
$$

\nit
It remains only to decompose the fields and momenta into longitudinal and
transversal
parts
$$
\ca_i=\ca_i^T-\frac{\pa_i}{\sqrt{\D}}\ca^L,\hspace{.5cm}\pa_i\ca_i^T=0,$$$$
P_i=P_i^T-\frac{\pa_i}{\D}P^L,\hspace{.5cm}\pa_iP_i^T=0,
$$
$$
[P^T_i(x),\ca^T_i(y)]=-i\d_{ij}\d^3(x-y),
$$
$$
[P^L(x),\ca^L(y)]=i\sqrt{\D}\d^3(x-y)
$$

\nit
and realize that the Hamiltonian is a sum of independent three oscillators
(the term in exponent in (\ref{WZW}) depends on $\ca^T$ only):
$$
H=\int d^3x \lp (\half P_i^T{}^2+ \half m^2 \ca_i^T{}^2) + (m^{-2}\half
P_i^L{}^2+
\half m^2 \ca_i^L{}^2)\rp.
$$

\nit
Therefore the wave function resulting from the Dirac quantization
is the wave function of three independent oscillators (Hermite polynomial
times the exponential dumping factor)\footnote{The argument
for the longitudinal part is $A^L\D^{-1}A^L$.} multiplied by the
term (\ref{WZW}). We will not discuss here the problem of
construction of the inner product.

The Gupta -- Bleuler procedure is straightforward. One takes the
holomorphic constraint to be $P_0-iQ_0$ and realizes that
the wave function as
$$
J_{GB}(Q_0,\ca_i) = e^{-\half \int Q_0^2}\J_D(\ca),
$$

\nit
where $\J_D$ is the wave function described above. We see therefore that
the Gupta -- Bleuler procedure merely re-introduces the variables which have
been cut out classically in the Dirac procedure and the addition
part of the wave function takes the form of the oscillator vacuum.
In view of the fact discussed in the previous section that  Gupta -- Bleuler
constraints can be identified with annihilation and creation operators
this result is very natural and indeed comprise the essence of the
Gupta -- Bleuler procedure.

To finish this section let me draw the reader's attention to the following
important fact. Contrary to the Dirac procedure where the physical states are
not normalizable (as the wave function does not depend on some variables) in
the
Gupta -- Bleuler procedure, as it was discussed above, the physical
wave function contains exponential dumping factor and the physical
states are perfectly normalizable.  This circumstances make it possible
to interpret the inner product of physical states directly in terms of
physical probabilities. Therefore, one avoids the neccesity of introducing
projectors on physical states, which is one of the major technical problems
of Dirac quantization\footnote{One could think about overcome this problem as
follows. Given Dirac (first class) constraints, add gauge fixings, polarize the
system and follow the Gupta -- Bleuler procedure. This proposal is certainly
worth further investigation, however in many theories finding the correct
gauge fixings is a very difficult problem  itself.}.

\sect{Path integral}

The path integral technique (for review see e.g. \ct{FS})
is nowadays the most popular way of
quantization. The simple reason behind it is that it serves the
most economic way of obtaining Feynman rules (i.e., the
description of perturbative sector of quantum theory) directly
from classical Lagrangian -- it amounts in
formal manipulations of the vacuum-to-vacuum
amplitudes
\be
Z[J] = \int \cd\m exp\{i(S_{cl} + J\f)\},             \lbl{pi}
\ee

\nit
where $S_{cl}$ is a classical action, and the sources $J$ are coupled to all
fields $\f$, and $\cd\m$ is an appropriate measure.
 Complementary to canonical
quantization in which the time evolution of wave function
is unitary by construction but some global symmetries
of the theory are not obviously preserved, in path
integral all symmetries are manifest\footnote{As observed by Fujikawa
\ct{Fuji}, even though $S_{cl}$ is invariant with respect to all relevant
symmetries of the problem by construction, the measure, $\cd\m$ may change by
the non -- trivial
Jacobian. This is the way how the anomalies show-up in the path integral
language.}
and one needs to check  unitarity of the S-matrix\footnote{Obviously, this
holds true for
Lagrangian path integrals. The Hamiltonian path integral which I will
consider below has the same virtues as the canonical quantization procedure.}.

It is not easy if not impossible to derive the expression (\ref{pi})
directly from the canonical analysis in general case. What one can do
more or less rigorously is to express the matrix element of
the evolution operator in terms
of the path integral over the phase space for unconstrained
system, to wit
\bq
<q',t'|q,t> &=& <q'|e^{i(t-t')H}|q> \nonumber \\
&=& \int_{path}\prod_{t}dq\frac{dp}{2\p }exp\{i\int_{t}^{t'}dt
(p\dot{q} - H(p,q))\}.      \lbl{hpi}
\eq

\nit
Then one can try to integrate over momenta to arrive at the
formula (\ref{pi}). This is however possible only
if the  integral is Gaussian.

Let us pause here for a while to see how formula (\ref{hpi}) can be derived
and, what is more important, to understand what is the meaning of the symbols
that appear in it\footnote{This subject is extensively and deeply discussed in
the
book by Berezin and Shubin \ct{BS} and the reader can find there  lots of
details which we omit here.}.

Let us consider the probability amplitude $<\vec{q}',t'|\vec{q},t>$
(we use vectors over the
symbols to remind the multicomponent nature of $|\vec{q}>$, $|\vec{p}>$). Due
to the
principle of unitarity, the time interval $(t' - t)$ can be split into
$N~+~1$ subintervals, each of the length $\e$ and the first equation
in (\ref{hpi}) can be rewritten as\footnote{We assume that the Hamiltonian
$H$ is not explicitly time-dependent.}
\bq
&&<\vec{q}' , t'|\vec{q}, t> = \nonumber \\
&=& \int (\prod d\vec{q}_{i}) <\vec{q}'|e^{-i\e
H}|\vec{q}_{N}><\vec{q}_{N}|e^{-i\e H}|\vec{q}_{N-1}>
\cdots <\vec{q}_{1}|e^{-i\e H}|\vec{q}>.
\label{1}
\eq

\nit
This formula is still exact, but the presence of operators in the exponents
makes it not very useful. The idea is to replace the matrix elements of
operators by some other object, which is easy to calculate. To make this
crucial step, we need to introduce a notion of the classical symbol of
operator. First of all, we must resolve the ordering problem by defining an
expansion of any operator $\hat{A}$ into momentum and position operators
$\hat{p}$, $\hat{q}$, respectively (for a while we introduce hats here
to distinguish quantum-mechanical operators), to wit
\bd
\hat{A} = \sum \hat{q}^{i_{1}} \cdots \hat{q}^{i_{\a}}
\ca_{i_{1}\ldots i_{\a}j_{1}\ldots j_{\b}}\hat{p}_{j_{1}}\cdots\hat{p}_{j_{\b}}
\ed

\nit
Taking the position and momentum eigenstates, $|\vec{q}>$, $|\vec{p}>$,
$<\vec{q}|\vec{p}> =\\ \exp (i\vec{q}\cdot\vec{p})$, one can
easily find that the matrix element of $\hat{A}$ equals
\bd
\ca (\vec{p},\,\vec{q}) = <\vec{q}|\hat{A}|\vec{p}>e^{-i\vec{q}\cdot\vec{p}},
\ed

\nit
where,
\bd
\ca = \sum \ca_{i_{1}\ldots i_{\a}j_{1}\ldots j_{\b}}p_{j_{1}}\cdots p_{j_{\b}}
q^{i_{1}} \cdots q^{i_{\a}}.
\ed

\nit
Now we are in position to define the path integral, eq. (\ref{1}). We have,
\bqn
&&<\vec{q}_{i}|e^{i\hat{H}\e}|\vec{q}_{j}> = \\
&=& \int\frac{d\vec{p}}{2\p
h}<\vec{q}_{i}|e^{i\hat{H}\e}|\vec{p}><\vec{p}|\vec{q}_{j}> = \\
&=& \int\frac{d\vec{p}}{2\p
h}<\vec{q}_{i}|e^{i\ch\e}|\vec{p}><\vec{p}|\vec{q}_{j}>
e^{-i\vec{q}\cdot\vec{p}} + \co (\e^{2}).
\eqn

\nit
where $\ch$ is the symbol of the operator $\hat{H}$ and $\co (\e^{2})$ terms
reflect
the
difference between the symbol of exponent and exponent of the symbol.

Now, it is customary to take the limit $\e \rightarrow 0$, $N\e = t- t'$ and
denote the result as
\be
\int [dp][dq]e^{-i\int (p\dot{q} - \ch (p,q))dt} = \int
[dp][dq]e^{-iS_{cl}}.\lbl{i1}
\ee

\nit
Some important remarks are in order
\begin{enumerate}
\item The integration in (\ref{i1}) runs over the set of eigenvalues
of momentum and position operators in their eigenstates. This shows again
that the path integral makes use of the semiclassical and not classical
objects.
\item The $\dot{q}$ in the final formula is a shorthand notation for the
expression $\lim_{\e\rightarrow 0}\frac{q_{i+1} - q_{i}}{\e}$.
But, as it is well-known, the major contribution to the path integral comes
from the
trajectories, which are not differentiable. One should also carefully check
if the $\co (\e^{2})$ terms do not contribute to the path integral. One
may also understand this term as  reflection of inherited ambiguity in
definition of
the path integral which disappears if one sticks to one well -- defined
prescription
of how path integral is constructed. This subtle point is discussed in
\ct{BS}.
\item Even more important is the simple but often overlooked fact that
the terms in the exponents of path integral are {\em the symbols of operators
and {\em not} the classical objects}. These two may well coincide, but there
are important examples where they {\em do not}. Actually, the anomalous
theories
serve as the primary examples of such situation. The point is that if the
Poisson bracket appears in the path integral exponent, one should understand it
as the quantum mechanical matrix element of the corresponding commutator,
between
position and momentum eigenstates (it is easy to see that instead of momentum
and
position eigenstates one can take any other complete set of eigenstates as, for
example,
the particle number operator eigenstates (Fock basis)).
\end{enumerate}

Keeping the derivation and comments in mind, let us now turn to  discussion of
the path integral quantization of gauge theories. Below, we will
discuss only the phase space path integral which, contrary to
the Lagrangian path integral,
is directly linked to the most fundamental canonical quantization
technique\footnote{%
The problem with the Lagrangian path integral is that, in general, there is
little
control over theory being quantized, especially when problems like
anomalies arise.}.

\vspace{3ex}
For constrained systems, the theory
of phase space path integral was  developed by Faddeev \ct{Fad}
in the case of first class irreducible constraints  and
generalized by Senjanovic to incorporate the second-class ones
\ct{Sen}.
The main idea was simple to postulate a form of the path integral by
adding the delta functions of constraints and gauge fixings together with
appropriate determinants to the measure and integrate over
the unrestricted phase space of the problem. Then one checks
if the resulting expression
is the same as one obtained from (\ref{hpi})  for the integration
over the reduced phase-space
of the problem (i.e., when the constraints are solved by means of (classical)
gauge fixings prior to
construction of the path integral). The  expression (\ref{pi})
can be then obtained after integration over momenta as before.

In the middle seventies, in the remarkable series of papers \ct{BF}, \ct{BV},
\ct{FF}
(see also reviews \ct{h}, \ct{BFr}) the most general form of phase space
path integral was  obtained by means of extending phase space as to incorporate
ghost fields, which have had been introduced earlier  in Lagrangian formulation
of path integral by Faddeev and Popov \ct{FP}.
In this extended phase space the local gage symmetries of the problem are
replaced
by single global symmetry called the BRST symmetry. This symmetry was
discovered
in the middle 70's as a symmetry of Faddeev -- Popov path integral and it
was realized later that it plays important role both in
classical and quantum theory. Since our formulation of path integral
is essentially based on the BRST construction, let us pause for a moment to
review it.

The volume of this paper does not permit even a brief overview of the BRST
theory,
the reader should again refer to \ct{HT} for more detailed information. Let us
just
recall the most basic notions.

Given a set of (Hermitean)
first -- class constraint operators $\cg_{\a}$\footnote{For simplicity I assume
that
all constraints are bosonic, irreducible, and their algebra is not open
($f_{\a\b}^{\g}$
below are constants). The most general case is discussed in \ct{HT}.}
$$
[\cg_{\a},\cg_{\b}]=f_{\a\b}^{\g}\cg_{\g}
$$
one defines the nilpotent BRST operator\footnote{In quantum mechanical
framework there is of course the ordering problem which must be resolved.
Here and below, I assume that some ordering prescription (in our case normal
ordering)
is always in force. Otherwise, one should add some additional terms,
e.g., to make (\ref{BRSTcharge}) Hermitean.}
\be
Q=c^{\a}\cg_{\a} - \half c^{\a}c^{\b}f_{\a\b}^{\g}\p_{\g},\lbl{BRSTcharge}
\ee
$$
[c_{\a},p_{\b}]_+ = \d_{\a\b},
$$
\be
[Q,Q]_+=0.\lbl{nilpcond}
\ee

Since the BRST charge $Q$ is nilpotent, it provides a well-posed cohomology
problem. To proceed, we introduce a grading in the space of wave
functions over extended phase space as follows: We say that the `ghost number'
of the ghost $c$ is $+1$ and of the `ghost momentum' $-1$. We assume further
that the wave functions are (finite) polynomials in ghosts
and therefore, they can be decomposed into a sum of terms with definite ghost
number.
Acting on such space, the BRST operator is a cohomology operator, since it
raises the grading by one. We define the BRST (state) cohomology by
$\sfH^{(n)}=Ker\,Q|_n/Im\, Q|_{n-1}$ (the kernel of $Q$ acting on the space
with ghost number $n$ divided by the image of $Q$ acting on the
space with ghost number $n-1$).
Then it is obvious that the cohomology space with no ghosts is the physical
space of Dirac. Indeed, $Ker\, Q|_0=\bigcap_{\a}\, Ker\, \cg_{\a}$ and
$Im\, Q|_{-1}=\emptyset$. It is assumed that the generator of dynamics has zero
ghosts
number and commutes with the BRST charge, therefore, the cohomology
does not change during the time evolution  and if one assumes that the initial
state is physical,
it will evolve to a physical state.

It is not necessaryin general  that the BRST charge has the form
(\ref{BRSTcharge}). In fact,
this is the so--called minimal form of this operator. One can add any
additional sector
with trivial cohomologies, for example, the sector containing the Lagrange
multipliers
$\l^{\a}$ with their momenta $\p_{\a}$ and the corresponding ghosts sector
$\r_{\a}$, $\s_{\a}$.
The additional charge
$$
Q^{add}=\r_{\a}\p^{\a}
$$
has clearly the zero cohomology being the wave functions independent of $\l$
and the zero cohomology space of $Q+Q^{add}$ is $\\sfH_Q|_0\otimes
\{functions\;\;
independent\;\; of\;\; \l\}$. It turns out adding that such a BRST charge is
very convenient in formulation of the path integral. For more
detailed discussion,  see \ct{HT}.

There is one more important requirement concerning the BRST charge. In order
to assure automatic decoupling of closed states $|\j>=Q|\k>$ from
physical states, the BRST charge must be Hermitian. Indeed, then
$$
<phys|\j>=<phys|Q|\k>=<\k|Q|phys>=0.
$$
Moreover, for Hermitean $Q$, the closed states have zero norm. In other words,
the hermicity of the BRST charge with respect to some inner product
guarantees that the inner product can be consistently restricted to the
cohomologies.

The system of constraints $\{\cg_{\a}\}$ is, of course, not unique, as it can
be replaced by
any other system $\{\L^{\b}_{\a}\cg_{\b}\}$. One of the important properties
of the BRST formalism is that such  transformation is generated by a canonical
transformation in extended phase space and therefore it leaves the measure
of the path integral, which is a Liouville measure by construction, formally
invariant.

The proof of this fact consists of two steps. Since $\L^{\b}_{\a}$ is
invertible by definition,
$\det \L^{\b}_{\a} \neq 0$. If $\det \L^{\b}_{\a} < 0$, by making
the canonical transformation of ghosts $\c^1\rar -c^1$, $\p_1\rar-\p_1$,
and all other ghosts invariant, we get the BRST charge corresponding to
the system of constraints $\{-\cg_1, \cg_a\}$, ($a>1$). Therefore,
without loss of generality we can assume that   $\det \L^{\b}_{\a} > 0$.
Now, $\L^{\b}_{\a}$ can be obtained as a composition of infinitesimal
transformations
\be
\L^{\b}_{\a}=\d^{\b}_{\a}+\e^{\b}_{\a}\lbl{inftrans}.
\ee
These transformations are, in turn,  canonical transformations generated by
$\c^{\a}\e^{\b}_{\a}\p_{\b}$.

Let us turn to discussion of dynamics of the theory. The BRST analogue of the
fact
that constraints are preserved in time is that the BRST charge
is time-independent. Therefore, one looks for extension of the canonical
Hamiltonian $H_0$ which commutes with $Q$, to wit
\be
H=H_0+\ldots, \;\;\; [H,Q]=0.\lbl{unitham}
\ee
Such a generator of time evolution is called the unitarizing Hamiltonian.
It is easy to observe that the unitarizing Hamiltonian is not uniquely defined
by
eq.\ (ref{unitham}). Indeed, one could replace
\be
H\rar H+[K,Q]_+.\lbl{gaugeK}
\ee
In the formula above, the function $K$ which is assumed to have
the ghost number $-1$, is called the gauge fixing fermion. In fact,
it can be shown (see \ct{HT}) that different choices of $K$
correspond to different gauge fixings conditions.
\newline

Let me now summarize all the previous results and present the
step-by-step procedure of
construction of the Hamiltonian path integral which I will follow below:
\begin{enumerate}
\item In the first step one constructs an extended Hilbert space built over
the phase space which includes both physical and ghost variables.
\item Next, one constructs the BRST operator acting in this Hilbert space
and being nilpotent and Hermitean. The zero cohomology space of this operator
is assumed
to coincide with the space of physical states of Dirac quantization.
\item Given the BRST operator, one chooses the gauge fixing fermion $K$
and constructs the unitarizing Hamiltonian
operator (\ref{unitham}). This operator plays arole of the generator of
dynamics.
\item The unitarizing Hamiltonian is used to derive the path integral
following the formal procedure explained above. It is important that
any formal manipulations on the path integral can be made only {\em after}
the symbols of  operators are computed, i.e., when the objects under
discussion are already well defined functions on the classical phase
space of the problem.
\end{enumerate}

Let us employ this procedure in the case of Gupta--Bleuler constraints.
Our first problem is to find an appropriate BRST operator. This step
has been already performed in the paper \ct{ZJJJ}. It was shown in this paper
that one can define the so--called `second order BRST operator' $\O$
which is Hermitean by construction and whose zero cohomology coincides with
the Gupta--Bleuler physical states. This operator is constructed as follows
\be
\Omega = \ad{s} Q + \ad{Q} s + \gamma Z - 2 \ad{s} s \beta,
\label{Om}
\ee

\nit
where $(\ad{s},s)$ are second level commuting scalar ghosts and $(\gamma,
\beta)$ is a pair
of real conjugate Fermi-type second level ghost operators:
\bq
\left[ s, \ad{s} \right] = 0, \label{cs}\\
\nonumber   \\
\left\{ \gamma , \beta \right\} = 1,    \label{cga}
\eq

\be
Q = c^{A} \cg_{A} + \frac{1}{2} (-)^{A} \:  c^{A} c^{B} f_{BA}^{\:\:\:\:\:\:C}
\pi_{C} + \; ...\label{Qoperator}
\ee

\be
\ad{Q} = c^{\dagger A} \ad{\cg}_{A} + \frac{1}{2} (-)^{A} \:  \ad{\pi}_{C}
\label{barQoperator}
   \bar{f}^{C}_{\:\:BA} c^{\dagger B} c^{\dagger A} + ...,
\ee

\be
\begin{array}{l}
Q^{2} = \lp\ad{Q}\rp^{2} = 0, \\
    \\
\left\{ Q, \ad{Q} \right\} = 2 Z,  \\
    \\
\left[ Z, Q \right] = \left[ Z, \ad{Q} \right] = 0.
\end{array}
\label{Qalg}
\ee

\nit
All details of construction of the inner product in the Hilbert space of the
problem were presented in \ct{ZJJJ}. It should be only mentioned at this point
that the cohomology of $\O$ reduces to a sum of (dual in some sense)
cohomologies
of $Q$ and $\ad Q$.
The main  result of this paper  can be summarized in the following diagram
\be
\begin{array}{ccccccccc}
 & V_{0} & \stackrel{Q}{\longleftarrow} &
V_{1}
 & \stackrel{Q}{\longleftarrow} & ...& V_{2a} & \stackrel{Q}{\longleftarrow} &
0  \\
Q^{\dg}Q\downarrow &  & & & & &
 & & \\
 & V^{\ast}_{0} & \stackrel{Q^{\dg}}{\longrightarrow} &
V^{\ast}_{1} &  \stackrel{Q^{\dg}}{\longrightarrow} & ... & V^{\ast}_{2a} &
\stackrel{Q^{\dg}}{\longrightarrow}
 & 0.
\end{array}
\label{cohsec}
\ee

\nit
In this diagram
the space $V_{2a}$ is first level ghost free and corresponds to the
physical space of the Gupta--Bleuler procedure. Moreover,
the spaces $V_{i}$ and $V^{\ast}_{i}$ are adjoint with respect to the inner
product
of the Hilbert space of the problem.

In the next step, we construct the unitarizing Hamiltonian. Let us assume for
simplicity
that the canonical Hamiltonian $H_0$ commutes with both $Q$ and $\ad Q$. There
exists
a very natural choice of the gauge fixing fermion $K$ which is simple
and leads directly to the desired form of the path integral, namely $K =\b$.
Indeed,
for such choice the unitarizing
 Hamiltonian reads
\be
H= H_0+[\ad Q,Q]\lbl{GBunitham}
\ee

\nit
Let us observe, that in the formula above $\ad Q$ can be interpreted as a gauge
fixing fermion for the Gupta--Bleuler BRST operator $Q$, which is a very
natural choice
indeed.

At this point we can forget the second level ghost whatsoever. One can think
that they were integrated out and are no longer present in the construction.

As it was said before, it is convenient to extend the BRST operator
by adding an additional sector of variables. Thus, let
\be
Q\rar Q+\r_A\l^{\dagger A},\;\;\;\; \ad Q\rar \ad Q + \ad{\r}_A\l^A\lbl{redefQ}
\ee

At this point, one must further specify  properties of the operators which are
introduced
in definitions of $Q$ and $\ad Q$. First of all, we identify the variables
$c^{\dagger A}$ with the momenta of the ghosts $\r_A$ (and accordingly for
their conjugate variables). Next, we define $\l^A$ to be momenta
of $\l^{\dagger A}$: $[\l^A,\l^{\dagger A}]=-i$. Then it is easy to see that
the zero cohomology
of $Q$ is a tensor product of the  Gupta--Bleuler physical states and the
states
independent of the Lagrange multiplier $\l^{\dagger A}$ enforcing the
constraints $\cg_A$.

With these definitions in hand, we can further simplify formula
(\ref{GBunitham}) and the
form of path integral. However, here one should proceed with care. The reason
is
that there are examples of theories (like strings) which have anomalies in the
ghost sector. For such theories, some of commutators in (\ref{GBunitham})
may have anomalous terms and such terms, if present, could not be
omitted. However, for simplicity, we will ignore this possible complication
(One of the applications of the formalism presented here is to formulate
the path integral for string theory away of critical dimension. This problem
will
be discussed in the separate paper.). Keeping this remark in mind, we can
proceed
and just compute ghost commutators. For the sake of simplicity of the following
few formulas, I will assume that the constraints $\cg$ and $\ad{\cg}$ are in
the
abelian form -- the reader can easily convince himself that the final
formula (\ref{final}) below remains  the same in the general case.

Thus, we have
$$
Q= c^{A} \cg_{A}+ \r_A\l^{\dagger A}
$$
$$
\ad{Q} = c^{\dagger A} \ad{\cg}_{A}+\ad{\r}_A\l^A
$$

\nit
and the final form of the symbol of the unitarizing hamiltonian
to be inserted into the formula for the path integral
(\ref{i1}) reads
\be
\ch=\ch_0 + c^{\dagger A}[ \ad{\cg}_{A},\cg_B]_Q\, c^B +
\l^{\dagger A}\ad{\cg}_{A}+\l^A\cg_A +i\ad{\r}_A\r_A.\lbl{final}
\ee

In the formula above we introduced the notation $[\ast,\ast]_Q$ to denote
the symbol of the commutator. Clearly, for the theories with
anomalies this symbol {\em does not} coincide with the Poisson bracket
and contains the additional anomalous terms.

There is one more important group of terms in the $S_{cl}$ in (\ref{i1}),
namely the kinetic terms. Apart from the kinetic terms for
classical phase space variables we have (in agreement with our choice of
hermicity
properties of this sector)
\be
S_{kin} = \ad{\r}_A\dot{c}^A-c^{\dagger A}\dot{\r}_A +i\l^{\dagger
A}\dot{\l}^A.
\lbl{kinterms}
\ee

\nit
To transform the
formulas above to a more familiar form, we can perform a couple of (formal)
manipulations on the ghost
and $\l$ sector of the resulting path integral. First, we integrate
over $\r$ and $\ad{\r}$ and then
we make a (singular) scale transformation:
$$
\ad{\cg}\rar\e^{-1}\ad{\cg},\;\;
c^{\dagger A}\rar\e c^{\dagger A},\;\; \l^A\rar\e\l^A.
$$

This transformation does not change the measure of the path integral, since
there is
as many $\l$ as $c$ variables and they have opposite statistics. Now, we can
take the limit $\e\rar0$ to obtain the final form of the path integral
\be
S_{GB}=p\dot q - \ch_0
-c^{\dagger A}[ \ad{\cg}_{A},\cg_B]_Q\, c^B -
\l^{\dagger A}\ad{\cg}_{A}-\l^A\cg_A.\lbl{GBaction}
\ee
\be
Z=  \int [dp][dq][dc]d[\ad c][d\l]d\ad{\l}]e^{-iS_{GB}}\lbl{GBint}
\ee

As it was mentioned in Section 3, is usually quite hard to polarize constraints
and this is
one of the major problems of Gupta -- Bleuler procedure. The remarkable
fact concerning the path integral (\ref{GBint}) is that one does not need to do
that.
In fact, one can start with path integral with holomorphic `constraints'
and antiholomorphic `gauge fixings' and go all the way back to
real constraints.

As it was discussed in Section 3, one of tha basic properties of the
 Gupta--Bleuler construction is that there
exists a one -- to -- one linear relation between real and holomorphic
(antiholomorphic) constraints, to wit
$$
\cg_A =\L_A^{i}\cg_{i},\quad \ad{\cg}_A =\L_A^{\dagger i}\cg_{i}.\lbl{h-r}
$$

Then, we have
$$
(\l^{A}\ad{\cg}_{A}
+ \l^{\dagger A}\cg_{A})= (\l^{A}\L_{A}^{\dagger i}
+ \l^{\dagger A}\L_{A}^{i})\cg_{i}=
\l^{i}\cg_i.
$$

\nit
As far as the ghost term is concerned,
$$
c^{A}[\ad{\cg}_A, \cg_{B}]_{Q}c^{B}\rar c^{i}[\cg_i,\, \cg_j]_Qc^j.
$$

\nit
In the formula above, the terms  proportional to
$\cg_i$ were included into $\l^i\cg_i$. Now it is easy to observe
that the Jacobians of transformations
$\l^A, \, \l^{\dagger A}\rar \l^i$ and $c^A,\, c^{\dagger A}\rar c^i$ are the
same
but since these fields have opposite statistics, they cancel in the measure.
Thus, the path integral for anomalous theories can be written as
\be
Z = \int [d\m]e^{iS_{cl} + c^i\o_{ij}c^j}.\lbl{finalformula}
\ee

\nit
where $S_{cl}$ is the standard classical action and the effect of anomalies
reside in the
additional ghost term.

It is clear that the formula (\ref{finalformula})
can be also applied in the case of theories with classical second class
constraints. In this case this formula coincides with the formula
derived by Senjanowic \ct{Sen} and this fact can be treated as the first
check of the above formalism.

It is interesting to note that the term $\o_{ij}$ above is
known in the case of four dimensional chiral QCD and one can try to
use (\ref{finalformula}) as a departure point for discussion of the
perturbative quantum
theory of anomalous four dimensional chiral QCD.

To finish this chapter, let us observe that the formula (\ref{finalformula})
clearly differs from the standard path integral formula for gauge systems. In
the latter case, the path integral reads
\be
Z = \int [d\m]e^{iS_{cl} + c^i\{ \cf_i, \cg_j\}b^j}.\lbl{standardformula}
\ee
The differences are twofold. First, the formula (\ref{standardformula})
contains twice as many ghosts $(c,b)$ as (\ref{finalformula}). Secondly,
according to the standard formalism, in (\ref{standardformula}) one
introduces the gauge fixings $\cf$ for the gauge generators $\cg$. It is
obvious that such approach leads to inconsistencies since $\cg$ are
not generators of symmetries in the quantum theory if anomalies are present.

The result (\ref{finalformula}) can be also understood as follows. In the
presence of anomalies,
the gauge symmetries of the system is reduced by half (as it is clear from the
Gupta--Bleuler
picture) and the anomalous constraints start playing a role of their own
gauge fixings.

\sect{The chiral Schwinger model}
In this  chapter, I  demonstrate how the ideas developed
above work in practice. There is not hard to find an example for this
demonstration -- indeed the  chiral Schwinger model
\ct{SchMod} provide us with the theory which is both simple
(actually quantum mechanically solvable) and interesting as a
toy model for four dimensional chiral QED or QCD.

As an anomalous theory, the chiral Schwinger model has been
for long time considered to be quantum mechanically inconsistent.
Only in 1984 Jackiw and Rajaraman, \ct{JR}, discovered that this model
can be consistently quantized and showed that the physical
spectrum  consists of the massive and massless scalars,
with mass parameter, $m$, proportional to the electric charge of the
original theory, $e$. However, Jackiw and Rajaraman did not refer
explicitly to the model in its original fermionic formulation,
but rather they considered the model with bosonized fermions.
This formulation  has a virtue that there are no anomalies:
the anomalous commutator in fermionic formulation was replaced
after bosonization by
the classical second-class constraint.
The serious disadvantage of this approach is however that it introduces
an ambiguity to the theory, in fact the proportionality
coefficient relating mass to charge cannot be calculated.
Also the very concept of chiral bosonization is not very
well understood and in passing from fermionic to bosonic
formulation  some hand waving arguments are needed
(see for example  the nice papers of Harada \ct{Har}).
Below, we quantize the
chiral Schwinger model in its original fermionic formulation
and show that the resulting theory is solvable and unambiguous.

The action describing the chiral Schwinger model on the space
time of topology of $S^{1}\times R^{1}$   reads
\be
I = \frac{1}{2\p}\int d^{2}x(-\frac{1}{4}F_{\m\n}F^{\m\n}
+ \bar{\j}\g^{\m}[i\dd_{\m} +
eA_{\m} ](1 - \g^{5})\j ),
\ee

\nit
where the flat Minkowski metric $\h_{\m\n} = (-1, +1)$ and
$\g^{0} = \left( \begin{array}{lr}  0 &-1\\1 & 0 \end{array} \right)$,
$\g^{1} = \left( \begin{array}{ll}  0 & 1\\1 & 0 \end{array} \right)$,
$\g^{5} = \left( \begin{array}{rr}  -1 & 0\\0 & 1 \end{array} \right)$.
In terms of vector potential components $A_{\m} , \m = 0,1$ and the
chiral component of spinor field $\c$ the above action can be
rewritten as follows
\bq
I &=& \frac{1}{2\p}\int d^{2}x ( \half (\dd_{0}A_{1} -
     \dd_{1}A_{0})^{2} - i\ad{\c}(\dd_{0}\c + \dd_{1}\c ) \nonumber \\
 &-& e\ad{\c}\c(A_{0} + A_{1} )) \nonumber \\
 &=& L_{em} + L_{F} + L_{int}.
\eq

\nit
The form of Lagrangian suggests that the theory will be greatly simplified
if we use the light cone coordinates $x^{\pm} = x^{0} \pm x^{1}$. Indeed
in this coordinates the Lagrangian reads simply
\be
L(x^{+}) = \frac{1}{2\p}\int dx^{-} ( \half (\dd_{-}A_{+} -
     \dd_{+}A_{-})^{2} - i\ad{\c}\dd_{+}\c
 - e\ad{\c}\c A_{+} )
\ee

\nit
We immediately see that there are constraints in the problem, to wit
\be
P^{+} = \frac{\dd L}{\dd (\dd_{+}A_{+})} = 0.
\ee

Using the expression for momentum of $A_{-}$,
\be
P^{-} = \frac{\dd L}{\dd (\dd_{+}A_{-})} = -\frac{1}{2\p}(\dd_{-}A_{+} -
     \dd_{+}A_{-}),
\ee
we can easily calculate the canonical Hamiltonian
\be
H_{c} = \frac{1}{2\p}\int dx^{-}(\half P^{-}P^{-}  - A_{+}(\dd_{-}P^{-} -
e\ad{\c}\c )),
\label{sham}
\ee

\nit
which is indeed very simple. After quantizing in coordinate representation,
$P = -i\frac{\dd}{\dd A}$ the expression (\ref{sham}) becomes a
quantum Hamiltonian operator.
In the formula above we solved already the fermionic second-class
constraints i.e., the fundamental fermionic bracket is
\be
[\ad{\c}(x^{-}), \c (y^{-})]_{+} = \d (x^{-} - y^{-}),
\ee

\nit
as usual.
{}From the consistency condition
\be
[P^{+}, H] = 0,
\ee

\nit
we obtain the secondary constraints
\be
\cg (x^{-}) = \dd_{-}P^{-} - e\ad{\c}\c = 0,
\lbl{gauss}
\ee

\nit
which is nothing but the Gauss law.

To proceed let us expand all
fields present in the problem into Fourier modes:
\bq
A_{-} &=& \sum_{n} a_{n}e^{-inx^{-}}, \hspace{.3cm}\ad{a}_{n} = a_{-n}, n\geq
0, \nonumber \\
P^{-} &=& \sum_{n} \p_{n}e^{-inx^{-}}, \hspace{.3cm}\ad{\p}_{n} = \p_{-n},
n\geq 0,
\nonumber \\
A_{+} &=& \sum_{n} \a_{n}e^{inx^{-}}, \hspace{.3cm}\ad{\a}_{n} = \a_{-n}, n\geq
0.
\eq

\nit
We will expand the fermionic field in an unusual way, which is
however essential for later purposes
\be
\c(x^{-}) = \sum_{n} u_{n}e^{inx}exp(-e\sum_{m}\frac{1}{m}(a_{m}e^{-imx^{-}} -
\ad{a}_{m}e^{imx^{-}})),
\ee

\nit
and
\be
\ad{\c}(x^{-}) = \sum_{n}
\ad{u}_{n}e^{-inx}exp(e\sum_{m}\frac{1}{m}(a_{m}e^{-imx^{-}} -
\ad{a}_{m}e^{imx^{-}})),
\ee

\nit
The use of such expansion can be explained as follows. From purely
pragmatic point of view, this is the expansion which directly leads to
the desired final identification of the spectrum of the chiral Schwinger model.
On the other hand, this kind of expansion is to be expected from the following
physical
reasoning. It is well-known that because there are no propagating photons in
two dimensions,
the states of the model should consist of two fermions linked by the string (in
the
sense of the old string model of QCD). Then the quanta $u$ and $\ad u$ could be
interpreted
as a fermion with `half of the string'. Finally, it should be mention
that the expansion above leads to the so-called `kinematic normal ordering'
advocated in \ct{IST} and successfully employed to compute anomalous
commutators
in range of two- and four-dimensional theories (see \ct{DT} and references
therein).

{}From these expansions we see that the canonical anticommutational
relations for the fermionic modes holds:
\be
[u_{n}, \ad{u}_{m}]_{+} = \d_{m,n}.
\ee

\nit
Also we take
\be
[\p_{n}, \ad{a}_{m}] = [\ad{\p}_{n}, a_{m}] = -i\d_{n,m}. \label{ccra}
\ee

\nit
{}From the fact that $[P_{-}, \c ] = [P_{-}, \ad{\c}] = 0$
we find that

\be
[\ad{\p}_{m}, u_{n}] = -e\frac{i}{m}u_{n+m},
\ee
\be
[\p_{m}, u_{n}] = e\frac{i}{m}u_{n-m},
\ee
\be
[\ad{\p}_{m}, \ad{u}_{n}] = e\frac{i}{m}\ad{u}_{n-m},
\ee
\be
[\p_{m}, \ad{u}_{n}] = -e\frac{i}{m}\ad{u}_{n+m}.
\ee

\nit
In terms of Fourier modes the Hamiltonian (\ref{sham}) reads
\be
H = \sum_{n>0} (\p_{n}\ad{\p}_{n} + \a_{n}\cg_{n} +\ad{\cg}_{n}\ad{\a}_{n}) +
\half\p_{0}^{2}.
\label{fsham}
\ee

\nit
Here $\cg_{n}$ and $\ad{\cg}_{n}$ are Fourier components of the
Gauss law constraint which are already in the Gupta-Bleuler form
\bq
\ad{\cg}_{n} &=& -in\ad{\p}_{n} + e\sum_{l}\ad{u}_{l-n}u_{l}, \hspace {.2cm} n
> 0 \\
\cg_{n} &=& in\p_{n} + e\sum_{l}\ad{u}_{l+n}u_{l}, \hspace{.2cm} n > 0 \\
Q &=&  \sum_{l}\ad{u}_{l}u_{l} + q_{0}.
\eq

\nit
In the last equation we introduce the normalization constant $q_{0}$
for later purposes. Let us observe that the constraints $\cg , \ad{\cg}$
commute with $u_{n}$  and $\ad{u}_{n}$, which was a reason for introducing
the unusual expansion for $\c$ and $\ad{\c}$ above.

To proceed, us define the fermionic Fock space as follows. First we split the
operators
$u_{l}, \ad{u}_{l}$ with respect to the  vacuum state $|0>_{F}$,
\be
u_{n} = \left\{ \begin{array}{c}
b_{n}, \hspace{.2cm} n \geq 0 \\
\ad{d}_{-n}  \hspace{.2cm} n < 0
\end{array} \right.    \label{v1}
\ee

\be
\ad{u}_{n} = \left\{ \begin{array}{c}
\ad{b}_{n}, \hspace{.2cm} n \geq 0 \\
d_{-n}  \hspace{.2cm} n < 0
\end{array}  \right.    \label{v2}
\ee

\nit
and
\be
b_{n}|0>_{F} = d_{n}|0>_{F} = 0.
\ee

\nit

It is easy to see that the charge constraint
$Q$ is first-class even on quantum level as it
commutes with both $\cg_{n}$ and $\ad{\cg}_{n}$.
The algebra of other constraints is of the Gupta-Bleuler type,
\be
[\cg_{n}, \cg_{m}] = [\ad{\cg}_{n}, \ad{\cg}_{m}] = 0,
\ee
\be
[\cg_{n}, \ad{\cg}_{m}] = e^{2}n\d_{m,n}. \label{anomal}
\ee

\nit
These formulae are derived in appendix.
In the derivation we use operators  $\f_{n} = \sum_{l}\ad{u}_{l-n}u_{l}$ and
$\ad{\f}_{n} = \sum_{l}\ad{u}_{l+n}u_{l}$ which are
nothing but (anomalous) currents and which together
with $\p$ operators satisfy the algebra
\be
[\f_{n}, \ad{\f}_{m}] = n\d_{n,m}, \hspace {.5cm}
[\f_{n}, \f_{m}] = [\ad{\f}_{n}, \ad{\f}_{m}] = 0,
\ee

\be
[\p_{n}, \ad{\f}_{m}] = [\ad{\p}_{n}, \f_{m}] = 0,
\ee

\be
[\p_{n}, \f_{m}] = [\ad{\p}_{n}, \ad{\f}_{m}] = -ie\d_{n,m}.
\ee

\nit
We can easily find a representation of these commutational relations
and (\ref{ccra})
\bq
\f_{n} &=& n\frac{\dd}{\dd \ad{\f}_{n}} \\
\p_{n} &=& -i\frac{\dd}{\dd\ad{a}_{n}} + \frac{ie}{n}\ad{\f}_{n} -
\frac{ie^{2}}{2n}a_{n} \\
\ad{\p}_{n} &=& -i\frac{\dd}{\dd a_{n}} - ie\frac{\dd}{\dd\ad{\f}_{n}} +
\frac{ie^{2}}{2n}\ad{a}_{n}
\eq

\nit
We can also write down the representation for $\cg$ and $\ad{\cg}$:
\bq
\cg_{n} = n\frac{\dd}{\dd\ad{a}_{n}} + \frac{e^{2}}{2}a_{n} \\
\ad{\cg}_{n} = -n\frac{\dd}{\dd a_{n}} + \frac{e^{2}}{2}\ad{a}_{n}
\eq

\nit
Due to the anomaly the Gauss-law constraints are second-class.  Taking
$\cg_{n}$
to be a holomorphic one, which together with $Q$ describes the physical states,
and proceeding as in Section 4, we find easily the Gupta-Bleuler Hamiltonian
\be
H_{GB} = -\sum_{n>0}\frac{ie}{n}\f_{n}\p_{n} +
\half\p_{0}^{2} + constraints ,
\ee

\nit
which commutes with $\cg_{n}$ by construction. Since  $\cg_{n}= \p_{n} +
\frac{ie}{n}\ad{\f}_{n}$
the form of the Hamiltonian can be further simplified to give
\be
H_{GB} = -\sum_{n>0}\frac{ie}{n}\f_{n}\ad{f}_{n} +
\half\p_{0}^{2} + \g_{n}\cg_{n} + \g_{0}Q .
\ee
\label{schham}

\nit
Let us observe that the  Hamiltonian (\ref{schham})   and the holomorphic
constraints $\cg_{n}$ are almost identical to the corresponding ones for the
scalar field discussed before. The only difference is the $\half\p_{0}^{2}$
term
in the Schwinger model Hamiltonian  and the presence of $Q$ constraint.

Having constructed all the relevant operators let us turn to
revealing the physical Hilbert space and the dynamics of the theory.
 The $Q$ constraint can be solved easily by observing
that it commutes with $\f$ and $\ad{\f}$ and annihilates the fermionic vacuum
state once the constant $q_{0}$ is properly adjusted. Then the physical states
annihilated by $Q$ are of the form\footnote{The states presented below are
certainly
annihilated by $Q$. To show that the opposite is also true, i.e., that all
the states with this property are of the form of (6.41) is more difficult.
This type
of questions has been extensively analysed in the context of string theory.
A reader may consult e.g. \ct{AlvG} for details.}
\be
\J = \sum\j_{i_{1}\cdot i_{M}j_{1}\cdot j_{M}} (a , \ad{a}, a_{0})
(\f^{\dg})^{\g_{1}}_{i_{1}}\cdot(\f^{\dg})^{\g_{M}}_{i_{M}}|0>_{F},
\label{schphys}
\ee

\nit
where the range of summation  is finite. Let us now act with
the holomorphic constraints on the physical states. Since $\cg_{n}$ commute
with fermionic operators and annihilates the vacuum by construction,
 the condition $\cg_{n}|phys> = 0$ reduces to
\be
(n\frac{\dd}{\dd\ad{a}_{n}} + \frac{e^{2}}{2}a_{n})\j = 0,
\ee

\nit
from which it follows that
\be
\j = e^{-\frac{e^{2}}{2}\sum_{n>0}\frac{1}{n}a_{n}\ad{a}_{n}}\j_{0}(a, a_{0}).
\ee

\nit
Acting on physical states, the Hamiltonian (\ref{schham}) takes a form
\be
H = e^{2}\sum_{n>0}\frac{1}{n}\ad{\f}_{n}\frac{\dd}{\dd \ad{\f}_{n}} +
\half\p_{0}^{2}.
\ee

\nit
The first term in this operator is nothing but the Hamiltonian operator for a
massive scalar field with mass $m^{2} = \frac{e^{2}}{2}$. To discuss the second
term
let us observe first that the configuration space of $a_{0}$ is $[0,1]$
with endpoints identified. Indeed there is a remnant of the gauge symmetries
of the initial theory still present, given by $\d a_{0} = g^{-1}\dd g$,
$g = e^{inx}, n \in {\bf Z}$, by virtue of which one can transform any
$a_{0}$ into above mentioned interval. Therefore the $a_{0}$ part of the wave
function is $e^{2\p ia_{0}}$ and the action of $\p_{0}^{2}$ merely shifts
all the energy levels by the same amount and therefore is physically
irrelevant. Therefore we conclude that the quantum theory of Chiral Schwinger
Model with a charge $e$ is equivalent to the family of the theories
of massive scalar field, with mass $m^{2} = \frac{e^{2}}{2}$, parametrized by
the
eigenvalues of the $\p_{0}$ operator.

\sect{The Antianomaly}
The anomaly, in the language of  canonical quantization
is the appearance of an additional central term in the  algebra of
classically first-class constraints. It may
happen therefore that for the classical Poisson bracket
algebra of symmetry generators
\be
\left\{ \cg_{I}, \cg_{J} \right\} = -f_{IJ}^{\:\:\:\:K} \: \cg_{K},
\ee

\nit
the corresponding commutator algebra reads
\be
\left[ \cg_{I}, \cg_{J} \right] = if_{IJ}^{\:\:\:\:K} \: \cg_{K} + \o_{IJ},
\label{an}
\ee

\nit
where $\o$  is the anomaly. This means that the system which possessed a set
of local symmetries on classical level, is not gauge invariant quantum
mechanically.

The lesson we learned from the sects. 3 and 4 is that using the
Gupta-Bleuler procedure any theory with a general constraints commutator
algebra, (\ref{an}) can be consistently quantized. Even more, since
the procedure described above clearly gives the standard answer for systems
which
can be quantized according to Dirac and can be applied also in  more general
situations, we will consider it as our quantization tool.

Let us now return to the systems with anomalies. We showed above that due
to quantum corrections some of the classical gauge symmetries may be not
present in quantum theory.
However the opposite situation may occur as well:
given the classical algebra of the form
\be
\left\{ \cg_{I}, \cg_{J} \right\} = -f_{IJ}^{\:\:\:\:K} \: \cg_{K} + \o_{IJ},
\label{aac}
\ee

\nit
the quantum counterpart may be closed, to wit
\be
\left[ \cg_{I}, \cg_{J} \right] = if_{IJ}^{\:\:\:\:K} \: \cg_{K}.
\label{aaq}
\ee

\nit
I will call this effect antianomaly.
{}From the physical point of view antianomaly means that the quantum
system has larger set of symmetries then its classical counterpart.
It is even possible that we have a classical theory with no symmetries
at all, which is becoming a gauge theory after quantization!
The example  of such theory is presented below.

Let us remark at this point
that no standard technique of quantization could be applied in this
case, as all of them require the recognition of  the constraints
algebra of the theory
already at the classical level i.e., prior to quantization.
Indeed, when one attempts to quantize a theory described by the set
of constraints, which form the second class algebra  according to
the  Dirac prescription, one uses the Dirac bracket and removes
the constraints $\cg_{I}$ from the theory. There is therefore no
way of seeing the effective gauge symmetry of quantum theory,
described by the algebra (\ref{aaq}). One may also observe that
if the number of  constraints $\cg_{I}$ is $N$, and the dimension
of the  phase space of the problem is $2D$ (in the infinite-dimensional
case one should count per space-point), the theory resulting
from the Dirac quantization of (\ref{aac})
will describe a system with $D - \half N$
physical degrees of freedom, as compared with $D - N$ degrees
of freedom  described by (\ref{aaq}). These two theories are therefore
inequivalent. At this point one may argue that we are free to choose
any prescription, however the explicit model discussed below
shows that only the second approach is correct. Let us stress
two points. Firstly, the very discussion of the antianomaly
problem is only possible because,
having in hand the Gupta-Bleuler procedure described above,
 we are no longer afraid to
quantize a theory with second-class constraints unsolved.
Secondly, the problems of the Dirac quantization propagates
to the path integral formulation of the theory with
second-class constraints, which is clear from the
derivation of such path integral \ct{Sen}.

The most important message from the derivation above is that
in general there is no direct relation between the symmetries of
quantum theory and its classical counterpart. Therefore one
should not use any information concerning classical symmetries
in an analysis of quantum systems, before proving that they
are identical to the quantum gauge invariances.
\newline

It is not hard to find a physical system with antianomaly. Take the (non -
chiral)
Schwinger model i.e., the two-dimensional QED
coupled to a single (for simplicity) Dirac fermion. It is
well known that this theory is anomaly free. Now let us integrate out
(in the path integral formalism)
a left handed fermion. The resulting effective action
will be non -- local, but  the locality may be reestablished
by introducing an  additional (chiral) scalar field into the
theory\footnote{There
are several technical problems which must be solved in a course of this
procedure.
Not least it happens that the additional parameter appears, which reflect
the ambiguity in calculation of the functional determinant. This parameter
can be however fixed by demanding the gauge invariance of the
resulting theory, as will be shown later. Also an additional outside
input is required if one
wants to obtain the chiral scalar instead of scalar field, see \ct{Har}.}
\ct{JR}.
The theory formulated in this way is, of course,
completely equivalent to the initial
Schwinger model (up to the ambiguity mentioned in  the footnote) and
therefore must be anomaly free. The virtue of introducing the
scalar to replace the chiral fermion
(which is nothing but the bosonization of the theory)
is, however, that the quantum
mechanical anomaly has been translated into a classical
second-class constraint
and there is no gauge symmetry  on the classical level.
 Since, as said before, quantum mechanically the model has  a $U(1)$ gauge
symmetry the antianomaly effect must occur.

Let us present the Lagrangian of the Schwinger model with  partially
bosonized fermion field, in the light cone coordinates.
\bq
L(x^{+}) &=&  \int dx^{-} \{  \half (\dd_{+}A_{-} - \dd_{-}A_{+})^{2}
+ i\ad{\j}_{R}\dd_{+}\j_{R}    + 2e_{R}\sqrt{\p}A_{+}\ad{\j}_{R}\j_{R}
\nonumber \\
&+& \half(\dd_{+}\f)(\dd_{-}\f)
- \half  (\dd_{-}\f)^{2}
+ e_{L}(\dd_{+} - \dd_{-})\f A_{-}  \nonumber \\
&-& \half e_{L}^{2}A_{-}^{2} +
e_{L}^{2}aA_{+}A_{-}
\},       \label{l}
\eq

\nit
where $a$ is the above mentioned ambiguity parameter and
to be as general as possible we denoted
 right (left) charges by $e_{R}$ $(e_{L})$ respectively. It is easy to see
that  there is no gauge invariance in the theory defined by the Lagrangian
(\ref{l}). We will show however that the corresponding quantum theory
has a local $U(1)$ symmetry.

It is straightforward to calculate momenta $\P_{\F}
= \frac{\dd L}{\dd (\dd_{+} \F)}$
and then to construct the canonical  Hamiltonian
\bq
\ch &=&  \int dx^{-} \{  \half P_{-}^{2} - A_{+}\dd_{-}P_{-}
+    \half
 (\dd_{-}\f)^{2}
 - 2e_{R}\sqrt{\p}A_{+}\ad{\j}_{R}\j_{R}
\nonumber \\
&+& e_{L}\dd_{-}\f A_{-}  + \half e_{L}^{2}A_{-}^{2} -
e_{L}^{2}aA_{+}A_{-}
\},       \label{h}
\eq

\nit
and the constraints
\be
\p_{\j_{R}} + i\j_{R}  = 0,
\label{c1}
\ee

\be
\cc =  \p_{\f} - \half\dd_{-}\f - e_{L}A_{-} = 0
\label{c2}
\ee

\be
P_{+}  = 0,
\label{c3}
\ee

\be
\cg = -\dd_{-}P_{-}   - e_{L}^{2}aA_{-}
-2e_{R}\sqrt{\p}\ad{\j}_{R}\j_{R} = 0.
\label{c4}
\ee

\nit
In formulae above we denoted momenta associated with the electromagnetic field
$A_{\m}$ by $P_{\m}$.
The  first constraint (\ref{c1}) is the standard second-class
constraint for fermions,
reflecting the fact that the kinetic term for these fields is linear
in the time derivative. Since it does not change in course
of quantization we simply solve it by using the Dirac bracket.
Then we find the   fundamental
anticommutator
\be
[ \ad{\j}_{R}(x) , \j_{R}(y) ]_{+} = \d (x - y).
\ee

\nit
The  constraint (\ref{c3}) is clearly  first-class and reflects the fact
that the wave function should be $A_{+}$-independent. The remaining two
constraints, if treated classically form a second-class system, to wit
\bq
\left\{\cg (x), \cg (y) \right\} &=& -2e^{2}_{L}a\d '(x-y)        \label{b1}\\
\left\{\cc (x), \cc (y) \right\} &=&  -\d '(x-y)              \label{b2}\\
\left\{\cg (x), \cc (y) \right\} &=&  -e_{L}\d '(x-y)              \label{b3}
\eq

\nit
Before proceeding let us recast the constraints into  simpler  form.
First, by making the canonical transformation $\p_{\f} \rightarrow
\p_{\f} - e_{L}A_{-}, P_{-} \rightarrow P_{-} - e_{L}\f$ we turn the
constraint (\ref{c2}) into the standard chiral boson constraint
\be
\cc = \p_{\f} - \half\dd_{-}\f .                \label{f1}
\ee

\nit
Then we diagonalize the constraints system by taking a linear combination
of them,
\bq
\cg &\rightarrow& \cg - e_{L}\cc \nonumber \\
&=&  -\dd_{-}P_{-} - e_{L}(\p_{\f} +  \half\dd_{-}\f) - e_{L}^{2}aA_{-}
-2e_{R}\sqrt{\p}\ad{\j}_{R}\j_{R},  \label{f2}
\eq

\nit
such that the  algebra of new constraints reads
\bq
\left\{\cg (x), \cg (y) \right\} &=& e^{2}_{L}(1-2a)\d '(x-y),
\label{d1}\\
\left\{\cc (x), \cc (y) \right\} &=&  -\d '(x-y),              \label{d2}\\
\left\{\cg (x), \cc (y) \right\} &=&  0.              \label{d3}\\
\eq

\nit
The dynamics of the theory is governed by the primary Hamiltonian
\bq
\ch_{P} &=&  \int dx^{-} \{  \half (P_{-} + e_{L}\f)^{2}
+ \half  (\dd_{-}\f)^{2} + e_{L}\dd_{-}\f A_{-}  + \half e_{L}^{2}A_{-}^{2}
\nonumber \\
&+&  A_{+}\cg +  \l\cc
\},       \label{hp}
\eq

\nit
where $\l$ is an Lagrange multiplier.
In the evaluation of these formulae we took into account the fact
that the Poisson bracket of fermionic bilinears $\ad{\j}_{R}\j_{R}$ is
equal to zero.  This, however, changes in the quantum theory -- indeed
it is well known that the commutator of fermion bilinears is
\be
[ \ad{\j}_{R}\j_{R}(x), \ad{\j}_{R}\j_{R}(y) ] = \frac{i}{4\p}\d '(x-y)
\label{fam}
\ee

\nit
Taking this into account we see that the quantum commutator of
the `Gauss law' constraint $\cg$ reads
\be
\left[ \cg (x), \cg (y) \right] =  i\d '(x-y)[e^{2}_{L}(1-2a) + e^{2}_{R}].
\label{cc}
\ee

\nit
We can also  calculate the commutator of the constraint $\cg$ with the
primary Hamiltonian,
\bq
\left[ \cg (x), \ch_{P} \right] &=& \int dy
\{ie^{2}_{L}(a-1)(P_{-} + e_{L}\f)(x)\d (x-y) \nonumber \\
&+&
iA_{+}\d '(x-y)[e^{2}_{L}(1-2a) + e^{2}_{R}]\}.
\eq

\nit
We see therefore that the `Gauss law' constraint can be made first-class
and therefore a generator of gauge transformation if $a = 1$ and
$e^{2}_{L} = e^{2}_{R}$. For these values of parameters the model
described by the Lagrangian (\ref{l}) is equivalent to the standard
Schwinger model and, as we just have shown, exhibits the antianomaly
phenomenon.

We discovered above that the  parameter $a$ should take the value $1$.
Therefore the theory described by Lagrangian (\ref{l}) with $a=1$
and with no right fermions should be equivalent to the fermionic
theory  discussed in Section 6. To check this let us  look at the equation
of motion following from this Lagrangian.

Assuming  appropriate behaviour of all fields at spatial infinity,
from the $\f$ field equation  (we drop the subscript $L$)
\be
(\der - \dir )(\dir\f + eA_{-}) = 0,
\ee

\nit
we learn that
\be
\dd_{-}\f = -eA_{-}.
\ee

\nit
Substituting this result into the $A_{-}$ equation of motion we see
that the combination of fields $\r = \frac{1}{e}\dd_{+}\f + A_{+}$ is
a massive scalar field:
\be
\dd_{+}\dd_{-}\r + e^{2}\r = 0,   \lbl{mass}
\ee

\nit
with mass $m^{2} = e^{2}$. Taking these two results into
account the equation for $A_{+}$   reads
\be
\dir^{2}\r + e\dir\f = 0.
\ee

\nit
Acting on the right-hand-side of this equation
by $\der$ and using (\ref{mass}) we see that
\be
\dir (e\r - \dd_{+}\f) = \dir A_{+} = 0,
\ee

\nit
which is the equation defining a chiral  scalar field.
Therefore the spectrum of the left chiral part the model
consists of one massive and one massless chiral scalar field
and agrees with the spectrum of the chiral Schwinger model
with left handed fermions described
above.

Let me conclude this section  with some  remarks. The model
described and solved above is of course artificial and its
sole role was to give an example of the antianomaly
phenomenon. However, as an extra bonus it provided us with a
way of deriving the actual value of  $a$ parameter.  It
remains still an open question whether there exist
less trivial (e.g. four dimensional) theories with antianomalies.
The constraints structure of such theories is of course
pretty easy to guess: for example one can take $\cg$ in (11)
to include a fermionic chiral current and then to arrange
$\o_{IJ}$ to be  $-i$ times the chiral anomaly. The hard part
however would be to construct an appropriate Hamiltonian,
leading to unitary, Lorentz invariant and hopefully
renormalizable quantum theory.
Finally, let me stress that antianomalies have nothing to
do with anomaly cancellation, which  is an arranging for
such a field content of the theory that the classical and
quantum symmetry groups are identical.

\sect{Conclusions}

Let me briefly summarize the contents of this paper. In the first
three sections I derived the canonical Gupta -- Bleuler approach to
quantizing of the anomalous theories. I presented a `cook book'
describing how one should proceed to quantize such a theory from the scratch.
In Section 6 I used this procedure to quantize the 2 -- dimensional Schwinger
model and in Section 7 (among other things) I showed that the result coincides
with the well-- known
standard one obtained in the past.

In my opinion, however, the most important result of this paper is the
derivation of
additional terms present in the Hamiltonian path integral formulations of
anomalous
theories. It may happen that the path integral in the form presented in this
section
can be applied to consistent quantization of anomalous theories in the physical
dimension $d = 4$ like chiral QED or QCD. It would be interesting to see what
the
predictions of such theories are, for example if the anomalies give rise
to mass generation, as they do in two dimensions, and if some realistic models
describing elementary particle physics can be generated in the framework of
such theories. Work in this direction is in progress.

\begin{center}
{\bf\large Acknowledgement}
\end{center}

It was a number of my friends and colleagues with whom I discussed the subject
of this
paper. First of all, I would like to thank Jan--Willem van Holten for long
discussions about path integrals. In fact, the presentation at the beginning
of Section 5 follows closely his unpublished draft of the book on quantum
field theory. I would like to thank W. Kalau for discussions on Gupta --
Bleuler
constraints and to Z. Hasiewicz and Z. Jaskolski for patiently explaining
me subtle point of BFV formalism. I benefited a lot from discussions
with M. Blau, J. Lukierski and P. van Nieuwenhuizen.

I would also like to thank the anonymous referee for his comments which
helped me to gradually improve this paper.

\sect{Appendix}

\renewcommand{\theequation}{A.\arabic{equation}}
In this appendix we collect some results of more technical nature
which are used in the main text.

To start, let us denote for integer $n$ $\f_{n} = \sum_{l}\ad{u}_{l-n}u_{l}$
and $\ad{\f}_{n} = \sum_{l}\ad{u}_{l+n}u_{l}$. Let us now calculate the
commutator
algebra of these objects:
\bq
&&[\f_{n}, \ad{\f}_{m}] = \nonumber \\
&=&\sum_{k,l} (\ad{u}_{l-n}u_{k}\d_{l,k+m} - \ad{u}_{l+m}u_{l}\d_{l-n,k}
)\nonumber \\
&=&\sum_{l} (\ad{u}_{l-n}u_{l-m} - \ad{u}_{l+m}u_{l+n} ) \nonumber
\eq

\nit
The sum in the last line of above expression is not well defined and requires
regularization. We do that by taking the range of summation in definition of
$\f_{n}$ and $\ad{\f}_{n}$ to be finite and range from $-L$ to $L$
where $L$ is large enough. Using this regularization, the last line
above becomes
\be
\sum_{-L}^{L}(\ad{u}_{l-n}u_{l-m} - \ad{u}_{l+m}u_{l+n} ). \lbl{funda}
\ee

\nit
If $n \neq m$ one can shift the summation range
in the second term to see that all terms can
cancel when $L$ goes to infinity and the sum is zero. For $n=m$ however
we have products
of non -- anticommuting operators and more care is needed.
Acting on states with finite number of fermions, such that only states
with $\ad{b}_{n}$, $\ad{d}_{m}$, $L > n,m$ are present, we rearrange
both sums to the normal ordered form and see therefore that
\be
\sum_{-L}^{L}(\ad{u}_{l-n}u_{l-m} - \ad{u}_{l+m}u_{l+n} ) = n\d_{n,m}.
\lbl{fundb}
\ee

\nit
We stress here that the appearance of the non -- trivial right-hand-side in
(\ref{fundb}) is a purely quantum effect related to the
ordering problem, as opposite to the classical theory (in particular
the Poisson bracket of $\f$ and $\ad{\f}$ vanishes). It should
be also stressed that this formula gives unambiguous result only
on the Hilbert space consisting of finite number of fermionic
excitations, for infinite number of fermions it again becomes
ambiguous. However, we extend our result also to the
wave functions with infinite number of fermionic states,
which will be needed in description of physical states.
Actually the non -- vanishing of the
RHS is responsible for the anomalous term in the algebra of
classically first-class constraints $\cg_{n}$ and $\ad{\cg}_{n}$, as will
be seen below.

In analogous fashion one can prove that
\be
[\f_{n}, \f_{m}] = [\ad{\f}_{n}, \ad{\f}_{m}] = 0,
\ee

\be
[\ad{\p_{n}}, \f_{m}] = [\p_{n}, \ad{\f}_{m}] = 0,
\ee

\be
[\ad{\p_{n}}, \ad{\f_{m}}] = [\p_{n}, \f_{m}] = -i\d_{n,m}.
\ee

\nit
Now we are in position to calculate the algebra of constraints:
\be
[\cg_{n}, \cg_{m}] = 0,  \hspace{.2cm}  [\ad{\cg_{n}}, \ad{\cg_{m}}] = 0.
\ee

\nit
It is easy to see however that
\bq
&&[\cg_{n}, \ad{\cg}_{m}] = \nonumber \\
&=& [in\p_{n}, e\f_{m}] + [im\p_{m}, e\ad{\f_{n}}]
+ e^{2}[\ad{\f}_{n}, \f_{m}] = ne^{2}\d{n,m}.
\eq
\nit
which is the anomaly of the Gauss law constraint.

\end{document}